\documentclass[sigplan,twocolumn]{acmart}
\usepackage{multirow}
\usepackage{wasysym}
\usepackage{microtype}
\usepackage{enumitem}
\usepackage{tikz}
\usepackage{graphicx}
\usepackage{booktabs}          
\usepackage{algorithm}         
\usepackage{algpseudocode}     
\usepackage{amsmath}           
\renewcommand\footnotetextcopyrightpermission[1]{}
\setcopyright{none}
\AtBeginDocument{%
  }

\newcommand{\Name}{MTRDMA}

\begin{document}

\title{\Name{}: Lightweight User-Space Control-Plane Isolation for Multi-Tenant RDMA}

\author{Hongyu Wang}
\email{why2021@hnu.edu.cn}
\affiliation{%
  \institution{Hunan University}
  \city{Changsha}
  \country{China}
}

\author{Peihao Huang}
\email{hph101001@hnu.edu.cn}
\affiliation{%
  \institution{Hunan University}
  \city{Changsha}
  \country{China}
}

\author{Bangbo Liang}
\email{lbb@hnu.edu.cn}
\affiliation{%
  \institution{Hunan University}
  \city{Changsha}
  \country{China}
}

\author{Lei Wang}
\email{jordanwanglei@hnu.edu.cn}
\affiliation{%
  \institution{Hunan University}
  \city{Changsha}
  \country{China}
}

\author{Guo Chen}
\authornote{Corresponding author.}
\email{guochen@hnu.edu.cn}
\affiliation{%
  \institution{Hunan University}
  \city{Changsha}
  \country{China}
}

\renewcommand{\shortauthors}{Wang et al.}

\begin{abstract}
In managed multi-tenant clouds, RDMA's lack of control-plane (CP) resource isolation lets a noisy tenant exhaust shared NIC microarchitectural caches, degrading victim throughput by up to 96.4\%. Data-path schedulers leave CP resources unprotected, and hardware approaches require FPGAs or SmartNICs rather than the commodity RNICs operators already deploy.

We present MTRDMA, a lightweight user-space CP isolation mechanism for commodity RNICs, applicable to managed runtimes such as containers and sandboxed processes. MTRDMA intercepts library-level CP operations to enforce hierarchical quotas while preserving the zero-copy data path, using atomic shared-memory quota state with per-process cached admission (${\sim}$0.13\,$\mu$s), a hybrid garbage collector that reclaims orphaned quota within seconds, and rdma-cgroup caps with ledger--kernel reconciliation to bound and detect residual tampering. On ConnectX-6 RNICs, MTRDMA cuts victim degradation under Queue Pair (QP) cache exhaustion from 94.5\% to 0.08\%, extends protection to Memory Region (MR) cache exhaustion, and scales to 256 concurrent tenants with negligible data-plane overhead.
\end{abstract}


\settopmatter{printfolios=true}
\pagestyle{plain} 
\maketitle

\section{Introduction}
\label{sec:intro}

Remote Direct Memory Access (RDMA) provides microsecond-scale latency and kernel-bypass data transfer for high-performance distributed systems~\cite{kvrdma,FaRM,Onesiderdma,guo2016rdma,cloudmeetrdma}. To improve utilization, operators consolidate workloads from multiple tenants onto hosts that share a single RDMA NIC (RNIC)~\cite{FreeFlow,cloudmeetrdma,SwiftRDMA,metascale,DockRDMA,predictableDCN,ballani2013chatty,End_to_end,shareDCN,serveless,deathstarbench}. In \emph{managed runtimes}---containers and sandboxed processes---the operator controls the platform and tenants bring arbitrary binaries that may misbehave~\cite{FreeFlow,Justitia}. This paper addresses control-plane (CP) resource isolation between such tenants on commodity RNICs.

RDMA's kernel-bypass design splits the NIC into two paths. The data path (DP) bypasses the kernel entirely: applications post work requests and ring doorbell registers directly. The CP---Queue Pair (QP) creation, Memory Region (MR) registration---enters the kernel through the uverbs interface, but the kernel treats these commands as opaque pass-throughs: it records that a process created a QP but not how much of the RNIC's finite on-chip SRAM (QP, MR, and WQE caches) each tenant consumes. Nothing therefore stops one tenant from exhausting these caches. Reproducing an attack from Kong et al.~\cite{understandrdma}, after a noisy neighbor creates 100 QPs a victim's throughput drops from 91.5 to 3.3\,Gbps (96.4\%) on an unsaturated link (\S\ref{sec:cache_attacks})---cache eviction, not bandwidth; the effect extends to MR and WQE caches~\cite{understandrdma,Collie}.

A solution in this setting must meet four deployment requirements: it must not require application modification or recompilation (R1), because tenants bring arbitrary binaries; it must not modify the kernel, drivers, or firmware (R2), because production fleets run vendor-validated kernels whose changes must pass a long revalidation cycle; it must run on commodity RNICs (R3), ruling out FPGA or SmartNIC dependencies; and it must leave the DP untouched (R4), since kernel bypass is the reason RDMA is deployed.

On the standard path, every QP or MR creation walks the same route: the application calls \texttt{libibverbs}, which issues a uverbs system call, and the driver then commits the state to the RNIC. Admission must sit before the commit: committed contexts are non-preemptable, and the harm---a victim's eviction from on-chip SRAM---happens at commit time. R1--R4 themselves veto the application, the kernel, and NIC firmware or hardware, confining the admission point to unprivileged user space. The kernel case is the subtle one: the stock kernel exposes no RDMA-object hook that authorizes individual creations, so kernel-path enforcement---a kernel module, eBPF~\cite{mcanne1993bpf}, or a custom LSM~\cite{wright2002lsm}---means kernel changes.

The remaining candidates are user-space mechanisms on this path (\S\ref{sec:existing_limits} examines each family), and two further constraints reduce them to one. DP schedulers~\cite{PeRF,Justitia} see traffic only after creation has committed, too late to reject it. A supervisor at the system-call boundary---a privileged daemon or a per-container watcher---also sits before the commit, but its per-creation IPC round trip violates the low-latency admission requirement (\S\ref{sec:cache_attacks}; quantified in \S\ref{sec:existing_limits}). The one remaining spot is the \texttt{libibverbs} boundary, the last point on the path before the kernel. Interposing there meets all four requirements at once: applications run unmodified (R1), the kernel, drivers, and NIC stay untouched (R2, R3), and the DP is not involved (R4). This boundary sees only the standard path; \S\ref{sec:threatmodel} describes how the platform closes the rest---mandating preload, guarding the uverbs device node, and capping committed objects with rdma-cgroup---so library-level admission and platform-level bounding together cover all creation paths.

But this choice has a built-in weakness. Conventional resource control polices from above: a kernel or a hypervisor runs at a higher privilege than the processes it constrains, so the constrained party cannot tamper with the mechanism. The \texttt{libibverbs} boundary has no such advantage---the interposition layer runs inside the tenant's own process, at the tenant's privilege level, and a tenant can simply skip or overwrite a check that lives inside its own address space.

\Name{} therefore splits enforcement in two. A thin interposer inside \texttt{libibverbs} only catches each CP call before it commits, consults the quota, and lets the call through or fails it; it holds no authoritative state. All authoritative quota state lives in a separate, operator-launched user-space Resource Manager that stays off the per-operation path, so admission needs no IPC and no single daemon serializes the tenants. Yet this equal privilege creates two problems.

The first is the quota ledger (Challenge~1: an untrusted ledger). At equal privilege, the ledger must be tamper-resistant, fast under bursty creation, and exact under concurrent updates---a trilemma. A privileged arbiter keeps an unforgeable ledger but pays an IPC round trip per check and serializes every tenant through one process; counters in the tenant's own memory are fast but tenant-writable and cannot express tenant- or global-level quotas at all. \Name{}'s move is to stop trying to make the ledger unforgeable and instead make forgery pointless: it keeps authoritative counters in a daemon-initialized shared-memory segment, reads them through a per-process optimistic cache that admits in ${\sim}$0.13\,$\mu$s without IPC (\S\ref{sec:accounting}), and caps every tenant's real commitment at the kernel---above the tenant's privilege---with rdma-cgroup object limits. A tenant can rewrite the ledger, but not the kernel's cap; reconciliation turns any drift into a detectable divergence (\S\ref{sec:eval-bypass}).

The second is reclamation (Challenge~2: invisible reclamation). A process that exits cleanly destroys its QPs and MRs, and the ledger subtracts them; one killed by SIGKILL or a crash does not. The kernel eventually reclaims those orphaned hardware contexts, but it does so asynchronously and invisibly to user space, so the ledger keeps charging the dead tenant and its quota leaks away. Linux also recycles PIDs aggressively, so reclaiming by PID alone would eventually release a live process's resources as if it belonged to the dead one. \Name{} gives up on learning the exact death time and instead closes both leaks with cheap redundancy: \texttt{pidfd} events report deaths within seconds, a periodic scan catches the deaths \texttt{pidfd} misses when its file descriptors run out, and a process's recorded start time makes a reused PID unambiguously a new tenant (\S\ref{sec:gc-design}, \S\ref{sec:gc-evaluation}).

\Name{} realizes this design as a library-level interposition layer, the shared-memory ledger, and a resource-management daemon. Like any library-level mechanism, \Name{} is a resource-containment layer rather than a security boundary; \S\ref{sec:threatmodel} states the residual scope. On commodity ConnectX-6 RNICs, \Name{} reduces victim degradation under QP cache exhaustion from 94.5\% to 0.08\% and under MR cache exhaustion from 76.3\% to 0.1\%, adds ${\sim}$36\,$\mu$s to QP creation with no measurable data-plane bandwidth overhead, protects unmodified HERD~\cite{kvrdma}, an RDMA key-value store, and scales to 256 concurrent tenants (\S\ref{Evaluation}).
To our knowledge, \Name{} is the first system to enforce control-plane resource quotas for unmodified applications on commodity RNICs without kernel, driver, or firmware changes.

This paper makes the following contributions:
\begin{itemize}[leftmargin=1em, labelsep=0.2em]
    \item We formulate CP resource isolation for managed multi-tenant RDMA as a requirements problem and show that, together with pre-commit, low-latency admission, the requirements leave the \texttt{libibverbs} boundary as the only viable enforcement point---one running at the tenant's own privilege level (\S\ref{sec:cache_attacks}, \S\ref{sec:existing_limits}).
    \item We characterize what quota bookkeeping can guarantee at equal privilege---bounded transient overshoot under concurrency, second-scale reclamation of orphaned quota, and detection of residual tampering---and design the mechanisms achieving these bounds: atomic shared-memory admission and a hybrid garbage collector (\S\ref{Design}).
    \item We implement and evaluate \Name{} on commodity RNICs, reducing victim degradation under cache exhaustion from 94.5\% to 0.08\% with negligible overhead and scaling to 256 tenants (\S\ref{Evaluation}). We release \Name{} as open source at \url{https://github.com/mtrdma/mtrdma}.
\end{itemize}

\section{Background and Motivation}
\label{Background}

\subsection{RDMA Microarchitecture Resource Model}
\label{sec:rdma_model}

RDMA architectures separate resource management from data transmission~\cite{understandrdma,RPCs,RdmaDesignGuidlines,ScalableRdmaRPC}. The CP handles operations such as QP creation and MR registration through the uverbs interface: user space marshals command structures into kernel buffers, the driver programs the RNIC via PCIe, and hardware contexts are pinned in on-chip SRAM ($\sim$2\,MB on commodity ConnectX-6). The DP bypasses this entirely: applications write Doorbell registers directly, triggering Direct Memory Access (DMA) transfers without system calls~\cite{Userspace}.

\textbf{Kernel opacity at the microarchitecture level.} As introduced in \S\ref{sec:intro}, the kernel mediates CP commands through uverbs but treats them as opaque: it tracks that a process created a QP, not how much on-chip SRAM each tenant occupies, and it exposes no interface to query cache occupancy. The uverbs path consults no per-tenant resource counters before allowing creation.

\textbf{On-chip resource constraints.} Commodity RNICs manage three microarchitectural resources critical to performance: (1) the QP Cache, which stores QP context entries; (2) the MR Cache (Memory Translation Table (MTT)/Memory Protection Table (MPT)), which stores memory translation entries; and (3) the WQE Cache, which stores receive queue entries. These caches are finite and non-preemptable: once allocated, a context cannot be forcibly evicted by software. Hardware cache Quality of Service (QoS) mechanisms (e.g., Intel Cache Allocation Technology (CAT)~\cite{CacheQoS}) have addressed similar resource contention in CPU caches, but no equivalent exists for RNIC on-chip SRAM. When a tenant creates excessive QPs or MRs, the hardware evicts victim entries via standard cache replacement policies, causing severe performance degradation even when link bandwidth is abundant.

\subsection{Resource Contention as a Quota Problem}
\label{sec:cache_attacks}

To demonstrate how the missing CP protection manifests in practice, we replicate the QP Cache exhaustion attack\footnote{We adopt the term \emph{attack} from Kong et al. to describe resource exhaustion, but emphasize performance isolation rather than security.} from Kong et al.~\cite{understandrdma} on a 100\,Gbps ConnectX-6 link. A victim tenant sends 1\,MB messages over 4 QPs, holding steady at $\sim$91.5\,Gbps. A noisy neighbor then creates $5 \times 20 = 100$ QPs simultaneously, each pushing 512\,B messages, flooding the RNIC's on-chip QP Cache and evicting the victim's QP state.

\begin{figure}[!t]
    \centering
    \includegraphics[width=0.85\linewidth]{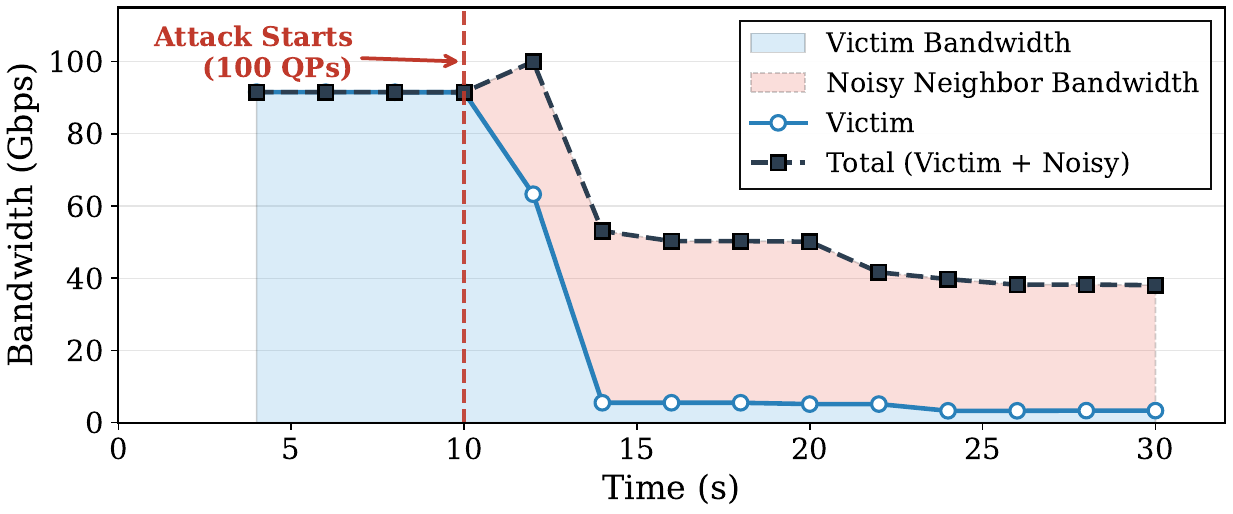}
    \caption{QP Cache exhaustion on a 100 Gbps ConnectX-6 link: the victim's bandwidth collapses from 91.5 to below 3.3 Gbps (96.4\% degradation).}
    \label{fig:qpc_attack_motivation}
\end{figure}

Figure~\ref{fig:qpc_attack_motivation} shows the timeline. Before the attack, the victim holds 91.5\,Gbps; once the 100 QPs start, its throughput drops to 63\,Gbps, then decays to 3.3\,Gbps---a 96.4\% degradation---while total link utilization settles at only $\sim$38\,Gbps, far below the 100\,Gbps capacity.

\textbf{Why bandwidth is not the bottleneck.} The degradation is not link saturation---the noisy neighbor uses 512\,B messages and total utilization is only 38\,Gbps---but cache eviction: the victim's 4 QPs are flushed from the on-chip QP Cache, and every subsequent packet forces a PCIe reload of QP context, destroying the latency and throughput advantages of kernel bypass. Similar phenomena extend to MR and WQE cache exhaustion (\S\ref{sec:mr-attack}) and have since been reproduced against containerized BlueField-3 tenants~\cite{kim2025noisy}.

\textbf{From contention to quota enforcement.} The attack succeeds because nothing prevents a tenant from unilaterally creating QPs until the on-chip cache is exhausted. The kernel's \texttt{rdma} cgroup tracks only coarse-grained, per-device object counts~\cite{kernelrdmacg,lwn2016rdmacg}, leaving a \textbf{control gap}: no mechanism stops a tenant from exceeding its fair share of CP resources before the damage occurs. What is needed is a \emph{hierarchical quota enforcement layer} that intercepts CP creation calls and rejects over-quota requests at the library boundary, before the kernel commits hardware state. Quota values themselves are deployment-time policy parameters (e.g., derived from Service Level Agreements (SLAs)); the system's role is to \emph{enforce} them with low latency and exact accounting, not to dynamically sense cache pressure. Admission must nevertheless be fast and low-overhead (R4): job startup bursts creations, while connections are long-lived~\cite{tsor2023}.

\subsection{Structural Blind Spots of Existing Mechanisms}
\label{sec:existing_limits}

No existing \emph{system} closes the CP resource creation gap on commodity RNICs (Table~\ref{tab:comparison}): each prior family has a structural blind spot; diagnostics (ReDMArk~\cite{ReDMArk}, Collie~\cite{Collie}) only detect, not enforce. The table then lists the two design choices for closing the gap on the standard path: the stock \texttt{rdma} cgroup cap, or in-process interposition at the \texttt{libibverbs} boundary.

\newcommand{\tabfull}{\tikz[baseline=-0.55ex]{\fill[black] (0,0) circle (0.9ex);}}
\newcommand{\tabempty}{\tikz[baseline=-0.55ex]{\draw[line width=0.65pt] (0,0) circle (0.9ex);}}
\newcommand{\tabhalf}{\tikz[baseline=-0.55ex]{\fill[black] (0,0) circle (0.9ex);\clip (0,0) circle (0.9ex);\fill[white] (0,-1ex) rectangle (1ex,1ex);}}
\begin{table}[t]
\centering
\caption{Comparison of isolation schemes against the deployment requirements R1--R4 (\S\ref{sec:intro}); the rightmost column is the CP-isolation goal.}
\label{tab:comparison}
\small
\setlength{\tabcolsep}{5pt}
\begin{tabular}{@{}lccccc@{}}
\toprule
\textbf{Scheme} & \textbf{R1} & \textbf{R2} & \textbf{R3} & \textbf{R4} & \textbf{CP} \\
\midrule
Harmonic~\cite{Harmonic}   & \tabfull & \tabfull & \tabempty & \tabfull & \tabfull \\
SCR~\cite{SCR}              & \tabfull & \tabfull & \tabhalf\textsuperscript{1} & \tabfull & \tabempty \\
Justitia~\cite{Justitia}   & \tabfull & \tabfull & \tabfull & \tabempty & \tabempty \\
PeRF~\cite{PeRF}            & \tabfull & \tabempty & \tabfull & \tabempty & \tabempty \\
\midrule
\texttt{rdma} cgroup (stock)~\cite{kernelrdmacg} & \tabfull & \tabfull & \tabfull & \tabfull & \tabhalf\textsuperscript{2} \\
\midrule
\textbf{\Name{} (Ours)}    & \tabfull & \tabfull & \tabfull & \tabfull & \tabfull \\
\bottomrule
\end{tabular}
\\[2pt]
\footnotesize
\tabfull = fully satisfied; \tabempty = not satisfied;
\tabhalf = partially satisfied.\\[2pt]
\textsuperscript{1} SCR requires NVIDIA BlueField-3 SmartNICs.\\
\textsuperscript{2} A possible design, but its aggregate cap admits no per-type quotas.
\end{table}

\textbf{Software DP schedulers: excluded by a timing gap.} Justitia~\cite{Justitia} and PeRF~\cite{PeRF} govern bandwidth sharing among \emph{existing} QPs---token-based pacing and preemption-based scheduling, respectively. Their structural blind spot is timing: they cannot intercept \texttt{ibv\_create\_qp} or other resource-creation verbs, so even under perfect DP fairness the victim starves because its QPs are evicted from the cache. Our experiments (\S\ref{sec:QPC-cache-attacks}) confirm that a QP Cache attack degrades the victim by 87.3\% under Justitia and 88.4\% under PeRF, necessitating \emph{pre-creation} interception (\S\ref{Design}).

\textbf{Hardware-dependent solutions: excluded by deployability.} Harmonic~\cite{Harmonic} (FPGA-based PCIe monitor) and SCR~\cite{SCR} (BlueField-3 SmartNICs) provide microarchitecture-aware isolation but require hardware other than the commodity ConnectX RNICs operators already deploy. The two also differ in kind: SCR is packet-granular \emph{data-plane} control and does not police CP creation, whereas Harmonic's PCIe monitor interposes before the driver commits state and can reject an over-quota creation pre-commit---hence their empty and full CP scores in Table~\ref{tab:comparison}, respectively. SR-IOV VF-per-tenant partitioning is natively available on the same adapters, but VF counts are limited, the partition is coarse, and prior work shows that SR-IOV and hardware traffic-class isolation do not isolate the RNIC's on-chip microarchitectural caches~\cite{understandrdma}. These deployability gaps necessitate a purely software mechanism on commodity RNICs.

\textbf{Kernel mechanisms: too coarse as shipped; finer means kernel changes (R2).} The first candidate is to lean on the kernel itself: every creation already traverses the kernel through uverbs, and the stock \texttt{rdma} cgroup caps per-device HCA handles and aggregate object counts per control group~\cite{kernelrdmacg,lwn2016rdmacg}. Its \texttt{rdma.max} interface defines only two nested keys, \texttt{hca\_handle} and \texttt{hca\_object}; the single \texttt{hca\_object} counter is shared by QPs, MRs, CQs, and PDs---no per-type split, no per-process tier within a tenant, and no reconciliation against operator policy. Closing these gaps in-kernel means changing the kernel and passing vendor revalidation, which R2 forbids. eBPF does not change this: the stock kernel's RDMA hooks govern only P\_Key access, subnet management, and security-context lifecycle, not per-object create authorization, and vetoing uverbs ioctls through the generic \texttt{file\_ioctl} hook would require a custom LSM---a kernel change either way. KRCORE~\cite{KRCORE22} demonstrates microsecond-scale control-plane elasticity but targets dynamic resource provisioning, not cross-tenant isolation, and is likewise a kernel module excluded by R2. Of upstream extensions, we have seen only byte-level MR accounting (\texttt{mr\_mem}) proposed under review~\cite{rdmacgmrmem}, which falls far short of covering the QP and MR cache attacks that object counts trigger; we measure how far this candidate gets us in \S\ref{sec:herd-protection}.

\textbf{User-space supervisors: excluded by admission latency.} The remaining family keeps enforcement in software but outside the tenant's process: a supervisor at the system-call boundary (e.g., seccomp user notification) or a privileged admission daemon, as deployed by container RDMA virtualization systems~\cite{FreeFlow}. These designs satisfy R1--R4; their structural cost is the per-operation round trip. Every creation crosses a process boundary (20--35\,$\mu$s per daemon arbitration versus ${\sim}$0.13\,$\mu$s for an in-process check, \S\ref{sec:accounting}), and a shared supervisor serializes admissions across tenants.

\textbf{Requirements for a deployable CP isolation layer.} In short, R1--R4 force the admission check into unprivileged user space, before commit, and in-process; the only position left is inside the tenant's own process---at equal privilege with the tenant it constrains, which is exactly what raises Challenge~1 (an untrusted ledger) and Challenge~2 (invisible reclamation) (\S\ref{sec:intro}).

\section{Design}
\label{Design}

\subsection{Overview and System Model}
\label{sec:principles}

\begin{figure}[t]
    \centering
    \includegraphics[width=0.42\textwidth]{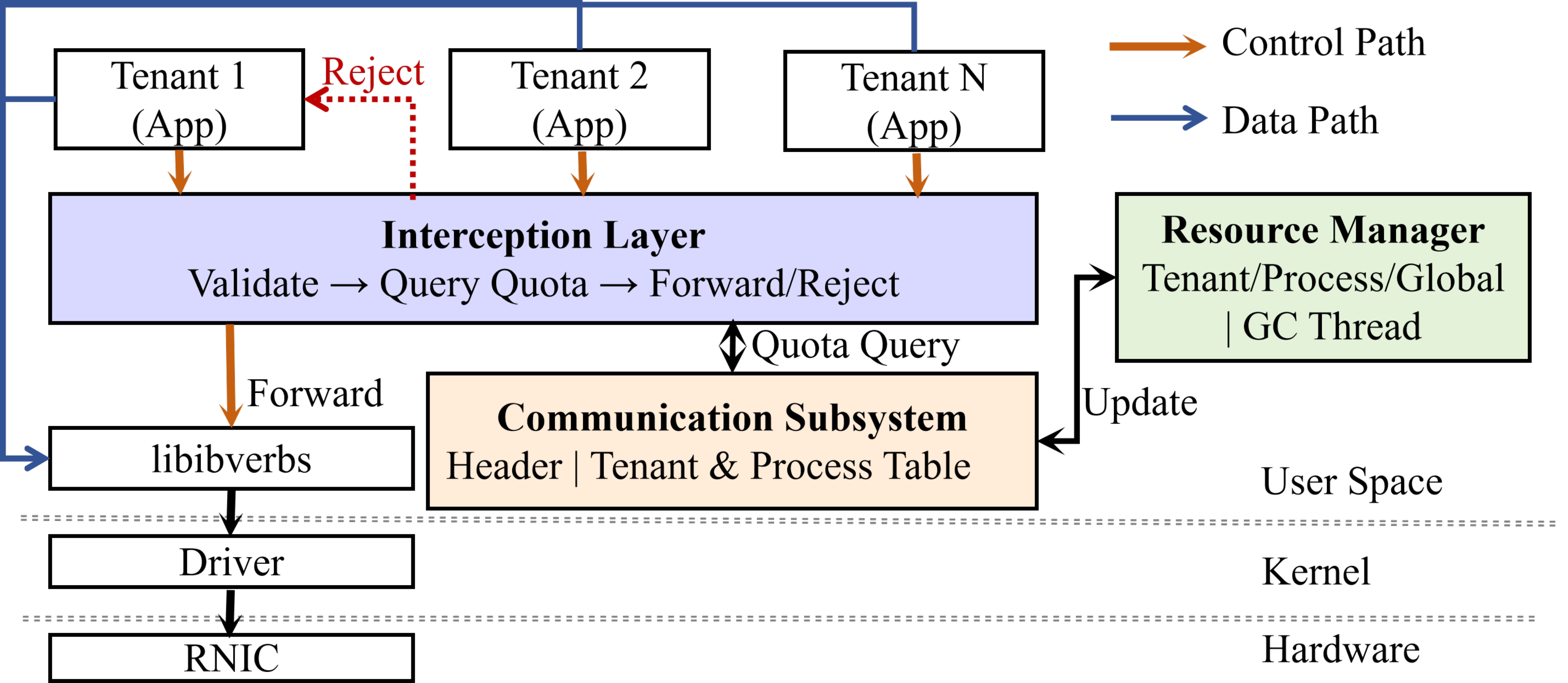}
    \caption{Three-layer architecture of \Name{}.}
    \label{fig:architecture}
\end{figure}

As established in \S\ref{sec:intro} and \S\ref{Background}, CP and DP are architecturally separable, enabling CP isolation via a lightweight user-space interception layer that leaves the DP hot path untouched. \Name{} closes the \emph{control gap} (\S\ref{sec:cache_attacks}) through the three-layer architecture shown in Figure~\ref{fig:architecture}.

The \textbf{Interception Layer}~\S\ref{sec:interposition} mediates application CP calls via library-level hooks. The \textbf{Communication Subsystem}~\S\ref{sec:accounting} provides low-latency, persistent coordination between the Interception Layer and the Resource Manager through POSIX shared memory. The \textbf{Resource Manager}~\S\ref{sec:gc-design} maintains authoritative quota state, enforces hierarchical admission control, and performs cross-lifecycle resource reclamation. Rejected requests return \texttt{EPERM} (operation not permitted) immediately.

\textbf{System Model.}
We consider a single physical node hosting tenants $\mathcal{T} = \{t_1, \dots, t_n\}$. Each tenant $t_i$ has a resource quota vector $\mathbf{Q}_i = (q_i^{\text{QP}}, q_i^{\text{MR}}, \dots)$, and the RNIC exposes finite hardware capacity $\mathbf{C} = (C^{\text{QP}}, C^{\text{MR}}, \dots)$. \Name{} enforces per-tenant quotas and global capacity constraints:
\begin{equation}
\forall t_i \in \mathcal{T}: R_i^{(r)} \leq q_i^{(r)}, \quad \sum_{t_i \in \mathcal{T}} R_i^{(r)} \leq C^{(r)},
\label{eq:quota}
\end{equation}
where $R_i^{(r)}$ denotes allocated resources of tenant $t_i$ for type $r$. We formalize \textbf{quota-enforced resource isolation} as follows: for any distinct tenants $t_i, t_j$, let $O_i$ denote CP operations issued by $t_i$ and $S_j$ the resource allocation state of $t_j$. \Name{} guarantees counting-level non-interference:
\begin{equation}
\forall O_i, \forall t_j \neq t_i: S_j \text{ before } O_i \equiv S_j \text{ after } O_i,
\label{eq:noninterference}
\end{equation}
because $t_i$'s cumulative allocation is capped at $\mathbf{Q}_i$ and cannot reduce the number of QPs, MRs, or other resources available to $t_j$. This is a \textbf{counting-level} non-interference property: it does \textbf{not} guarantee microarchitectural isolation---tenants' QPs may still share on-chip cache sets---an orthogonal concern requiring hardware-level mechanisms.

\textbf{Threat model and assumptions.}
\label{sec:threatmodel}
We scope \Name{} to resource exhaustion caused by \textbf{runaway, misconfigured, or malicious-but-contained tenants}: accidental resource leaks, autoscaling that exhausts on-chip caches, and deliberate CP flooding. A tenant controls code and data inside its container---root there, arbitrary syscalls, forking, and tampering with any reachable user-space state, including the shared-memory quota ledger, which is tenant-writable by design at equal privilege. Guarantees are accordingly stated as bounded damage plus detection, not as prevention of writes; \S\ref{sec:security-analysis} states the bound for each attack path, and \S\ref{sec:eval-bypass} measures it.

The trusted computing base comprises the host kernel, the cgroup subsystem, the POSIX shared-memory subsystem, and the Resource Management Daemon (including the guard launcher). The daemon and the guard run under an operator-controlled host identity, distinct from the unprivileged host users that run tenant processes; the binding table and other daemon-owned state are therefore read-only to tenants. We do not defend against (1)~side-channel attacks, a threat class orthogonal to resource contention; (2)~DP bandwidth saturation, delegated to hardware QoS (\S\ref{sec:hardware-qos}); (3)~full-guest-control VMs, whose passthrough control path never traverses host software and thus offers no enforcement point; or (4)~in-kernel RDMA consumers (NVMe-oF, IPoIB), platform-owned workloads outside the tenant quota model.

Three operational assumptions complete the model: enrollment tables are finite (the runtime's per-tenant process limit bounds identity and process-table slots, and a full table fails closed rather than evicting a neighbor---a fork bomb fills only its own tenant's slots); policy needs the daemon (during a daemon outage admissions continue under the last known quotas, fail-static, and new enrollment fails closed); and quotas sit below hardware capacity (protection presumes the global cap stays under the RNIC's on-chip threshold, 50--100~QPs per port on ConnectX-6~\cite{understandrdma}, since a higher cap cannot by itself prevent cache conflicts). \Name{} is thus a \textbf{resource containment} mechanism, not a security isolation primitive, adopting the managed-deployment model of prior user-space RDMA systems~\cite{FreeFlow,Justitia}. Within this model, every creation path to the RNIC is covered: the library path receives per-verb admission, and every path around the library is bounded by kernel-level caps and surfaced by reconciliation---\S\ref{sec:security-analysis} states each bound, and \S\ref{sec:eval-bypass} measures it. We present the three layers next.

\subsection{Interception Layer: CP-DP Orthogonality}
\label{sec:interposition}

The Interception Layer is the topmost tier of Figure~\ref{fig:architecture}, closing the \emph{timing gap} identified in \S\ref{sec:existing_limits} by intercepting CP calls before the kernel commits hardware state. It transparently intercepts library-level CP operations through dynamic linker interposition~\cite{linkers}, validates the tenant context, queries the Communication Subsystem for quota status, and makes an admission decision: approved requests forward to the original \texttt{libibverbs} implementation; rejected requests fail fast with \texttt{EPERM}, following classical defensive resource-isolation principles~\cite{verghese1998performance,shue2012performance,zhang2013cpi2}.

\textbf{Tenant identification.}
At library initialization, each process makes a one-time registration request over a Unix socket; the daemon authenticates the kernel-provided credentials, reads the process's RDMA cgroup path and \texttt{/proc/<pid>/stat} start time, and longest-prefix matches the path against the operator-maintained binding table. It then records $(\text{pid},\text{start\_time})\!\rightarrow\!\text{tid}$ in the shared Binding Table. A failed or unmatched registration fails closed: QP creation is rejected before entering \texttt{libibverbs} (\S\ref{sec:eval-bypass}). Forked children register as new process identities, and the start time prevents PID reuse from inheriting a stale binding.

Each intercepted CP operation looks up the caller's $(\text{pid},\text{start\_time})$ in the Binding Table and uses the resulting \texttt{tid} for the shared-memory accounting path. \texttt{RDMA\_TENANT\_ID} is therefore ignored for authorization and accounting---the bidirectional impersonation tests in \S\ref{sec:eval-bypass} show that changing it does not change which tenant is charged.

\textbf{CP-DP orthogonality.}
\Name{} scopes itself to CP admission only, leaving DP scheduling to existing mechanisms: the wrappers hook CP verbs while preserving the DP hot path and the exact InfiniBand Verbs (IB Verbs) Application Programming Interface (API) signatures. This orthogonality is validated with PeRF (App.~A); a scheduler that also hooks CP verbs would require symbol coordination, left to future work.

\textbf{Scope of interception.}
Our current implementation focuses on QP and MR verbs as representative CP operations that directly allocate RNIC on-chip resources (QP Cache and MR Cache, respectively). Empirical studies demonstrate that uncontrolled QP creation alone can degrade victim throughput by 96.4\% by flooding the QP Cache~\cite{understandrdma}, while excessive MR registration exhausts the on-chip MTT/MPT cache, causing 76.3\% degradation (\S\ref{sec:mr-attack}). The interception framework extends to additional CP verbs (e.g., Protection Domain (PD) allocation, Completion Queue (CQ) creation); we prioritize QP and MR as those directly correlated with on-chip SRAM exhaustion in commodity RNICs~\cite{understandrdma}. WQE cache protection is indirect: the firmware enforces a minimum receive-queue depth (32-deep QPs degrade a victim by 89.6\%), so capping QP count bounds the aggregate WQE footprint; a supplementary experiment (App.~B) evaluates a depth-level admission clamp, bounding the worst case at $Q \times \max(4{,}096, d_{\text{clamp}})$ entries. Within fixed quotas, varying queue depth to the largest creatable value (8\,192 send WRs), adding 50 CQs and 50 SRQs, or registering a 512\,MiB instead of 1\,MiB MR produced no measurable victim degradation: object counts, not per-object parameters, dominate the footprint on this device. Two coverage limits: 8\,192 is the largest depth the provider accepts, and the large-MR arm is a contiguous registration, not a multi-SGE scatter list.

\subsection{Communication Subsystem: Shared-Memory Quota Coordination}
\label{sec:accounting}

The Communication Subsystem keeps all quota state in daemon-initialized POSIX shared memory, which persists even if individual processes crash. Shared memory also keeps the synchronous quota check off the CP critical path: IPC sockets would introduce context-switch overhead (a daemon arbitration round trip measures $\sim$20--35\,$\mu$s vs.\ $\sim$0.13\,$\mu$s for the local cache), violating the low-latency admission requirement~\cite{FreeFlow}.

The state divides by writability. The \textbf{usage ledger}---Header (metadata, synchronization), \textbf{Tenant Table} (per-tenant usage counters), and \textbf{Process Table} (per-process consumption and affiliation)---is tenant-writable and serves the fast path: the optimistic cache below delivers \textbf{speed} (constant-cost checks), and lightweight in-memory spinlocks with atomic increments (\texttt{sync\_add\_and\_fetch}) deliver \textbf{concurrency} under mutually untrusted updates, without kernel-privileged arbitration. \textbf{Quotas} cannot live in that writable ledger: a tenant-writable quota field is a one-shot permanent denial---setting a victim's quota to zero rejects every later creation, and periodic reconciliation cannot tell an administrator's change from a forgery, so it cannot safely self-heal (\S\ref{sec:eval-bypass}). \Name{} therefore keeps per-tenant quotas in a separate daemon-owned \textbf{Policy Segment}: clients map it \texttt{O\_RDONLY} and the daemon owns the object, so a tenant cannot re-open it writable; the daemon re-asserts the segment on a one-second heartbeat, and each entry carries a sequence number so a reader never consumes a half-written update. Admission reads quotas only from this segment. The kernel-backed Binding Table is daemon-owned under the same read-only discipline, so a tenant cannot rewrite its affiliation while the accounting path stays writable and fast. This division---fast writable counters, read-only policy---answers the Challenge~1 trilemma (\S\ref{sec:intro}) at equal privilege. Table~\ref{tab:datastruct} summarizes the core data structures.

\begin{table}[t]
\centering
\small
\setlength{\tabcolsep}{3pt}
\renewcommand{\arraystretch}{0.95}
\caption{Core data structures maintained in the POSIX shared memory.}
\begin{tabular}{@{}p{1.7cm}p{3.1cm}p{2.7cm}@{}}
\toprule
\textbf{Structure} & \textbf{Key Fields} & \textbf{Purpose} \\
\midrule
Tenant Context & \texttt{tid}, \texttt{qp\_usage}, \texttt{mr\_usage} & Per-tenant usage counters (reconciliation-bounded) \\
Policy Segment & \texttt{tid}, \texttt{qp\_quota}, \texttt{mr\_quota}, \texttt{heartbeat} & Daemon-owned quotas; clients map read-only \\
Process Context & \texttt{pid}, \texttt{tid}, \texttt{qp\_local}, \texttt{mr\_local} & Per-process consumption and tenant affiliation \\
Binding Table & \texttt{pid}, \texttt{start\_time}, \texttt{tid}, \texttt{valid} & Kernel-backed identity; clients read-only \\
Global Snapshot & \texttt{total\_qp}, \texttt{total\_mr}, \texttt{timestamp} & Global guard; rebuilt from verified usage \\
\bottomrule
\end{tabular}
\label{tab:datastruct}
\end{table}

\begin{figure}[htbp]
    \centering
    \includegraphics[width=0.4\textwidth]{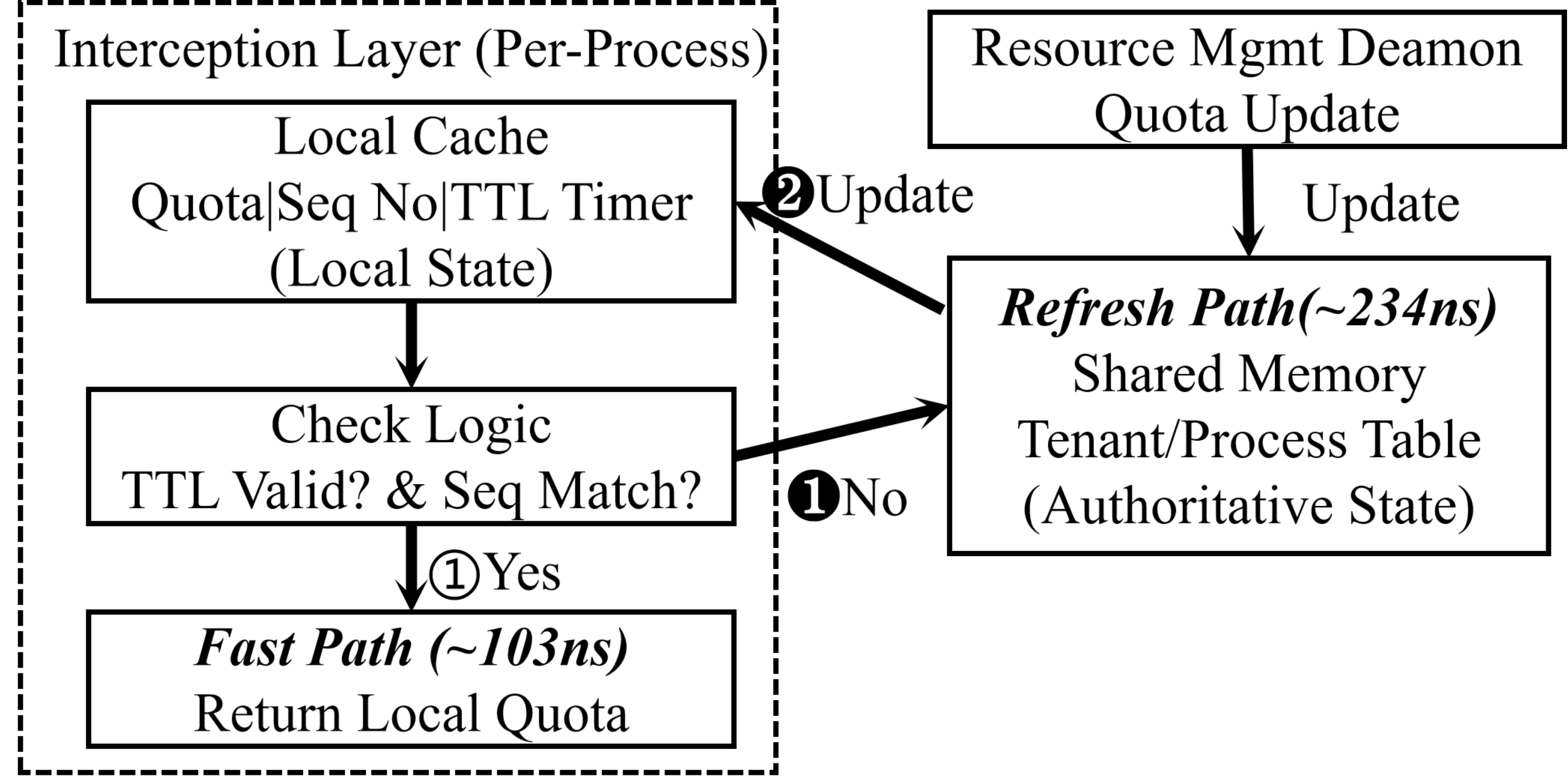}
    \caption{Two-path quota-check cache with TTL and sequence-number invalidation.}
    \label{fig:cache}
\end{figure}

\textbf{Fast-path admission with optimistic caching.}
To avoid shared-memory overhead under high tenant density, the Interception Layer maintains a per-process optimistic cache (256-bucket hash table, $\sim$2.5\,KB per process) storing hot quota metadata, a local sequence number, and a Time To Live (TTL) timer (Figure~\ref{fig:cache}). Each quota check first validates the local cache: if the TTL is valid and the local sequence number matches the entry sequence in the read-only Policy Segment, the request takes the fast path (${\sim}$0.13\,$\mu$s average over 300k checks); otherwise, it falls back to refreshing from shared memory (${\sim}$0.23\,$\mu$s) (supplementary App.~D). When the daemon updates a quota, it increments the entry sequence, and stale entries are detected lazily on the next access. This bounds inconsistency to milliseconds without violating safety: the authoritative shared-memory state remains consistent and the global hard cap prevents transient over-allocation.

\subsection{Resource Manager: Admission Control and Lifecycle Reclamation}
\label{sec:gc-design}

The Resource Manager answers Challenge~2 (invisible reclamation) (\S\ref{sec:intro}) by maintaining authoritative policy state, enforcing hierarchical admission control, and performing cross-lifecycle resource reclamation.

\textbf{Hierarchical quota enforcement.}
\Name{} enforces quotas at three levels as formalized in \S\ref{sec:principles}: per-process, per-tenant, and global; the per-process tier bounds intra-tenant fairness rather than acting as a security boundary, since a tenant that multiplies processes remains bounded by its tenant-level cap. The admission decision follows a strict hierarchical order: process limits first, then tenant limits, then global limits. If any check fails, the request returns \texttt{EPERM} immediately; no kernel calls or hardware operations are wasted. Algorithm~\ref{alg:admission} details the QP creation admission procedure; the same logic applies to MR registration with corresponding counters.

\begin{algorithm}[htbp]
\footnotesize
\caption{QP Creation Admission Control}
\label{alg:admission}
\begin{algorithmic}[1]
\Require Tenant ID $tid$, requested QP count $k$
\Ensure Boolean \textit{granted}, error code $err$
\State $p \gets \text{ProcessTable}[\text{pid}]$
\State $t \gets \text{TenantTable}[tid]$
\If{$p.\text{qp} + k > p.\text{qp\_limit}$}
    \State \Return \textbf{false}, \texttt{EPERM} \Comment{Process-level guard}
\EndIf
\If{$t.\text{qp} + k > t.\text{qp\_quota}$}
    \State \Return \textbf{false}, \texttt{EPERM} \Comment{Tenant-level guard}
\EndIf
\If{$\text{GlobalSnapshot}.\text{total\_qp} + k > C^{\text{QP}}$}
    \State \Return \textbf{false}, \texttt{EPERM} \Comment{System-level guard}
\EndIf
\State $\textit{success} \gets \text{ibv\_create\_qp}(\dots)$ \Comment{Delegate to real \texttt{libibverbs}}
\If{\textit{success}}
    \State $\text{AtomicInc}(p.\text{qp},\; t.\text{qp},\; \text{GlobalSnapshot}.\text{total\_qp})$
\EndIf
\State \Return \textit{success}, $0$
\end{algorithmic}
\end{algorithm}

\textbf{Resource lifecycle tracking.}
The Interception Layer maintains accounting through three paths: \textbf{creation} (validate the tenant context, forward compliant requests to the real \texttt{libibverbs}, and atomically increment tenant-, process-, and global counters); \textbf{destruction} (invoke the real verb first and decrement only on success, preventing undercounting from invalid handles); and \textbf{termination} (crashed applications leave orphans to the daemon's hybrid garbage collector below).

\textbf{Cross-lifecycle reclamation.}
RNIC contexts are non-preemptable: when a process receives SIGKILL, it cannot execute \texttt{ibv\_destroy\_qp}. While the kernel's ucontext cleanup eventually destroys orphaned hardware contexts, this reclamation is invisible to user-space quota counters. If \Name{}'s shared-memory state is not updated, the tenant's quota remains artificially exhausted even though the hardware has freed the resources---a \emph{counting-level leakage} (\S\ref{sec:intro}). We distinguish between hardware-level reclamation (performed by the kernel) and counting-level reclamation (performed by \Name{}). The hybrid garbage collector directly addresses the two root causes of Challenge~2: \emph{life-cycle asynchrony} and \emph{PID reuse}. It combines: \textbf{event-driven reclamation} via \texttt{pidfd}, achieving second-scale response; \textbf{periodic scanning} as a safety net for event loss under high concurrency; and \textbf{PID reuse protection} by recording each process's start time alongside its PID (the 64-bit \texttt{starttime} provides $2^{64}$ identities, making collision infeasible). Notification alone would silently miss deaths whenever \texttt{pidfd} slots are exhausted, and scanning alone would leak quota for a whole interval whenever an event is lost; the hybrid covers both.

Reclamation must also wait for the kernel, not only for the notification. A forked child inherits its parent's open device descriptor, and the kernel keeps the parent's QP objects alive while any inheritor holds it; zeroing the ledger on death alone would free quota that live kernel objects still consume. \Name{} therefore parks a dead identity in a pending state and releases quota only when the kernel confirms object release, gated conservatively on the identity's cgroup \texttt{rdma.current} count returning to baseline (per-PID netlink counts are unreliable: the kernel may omit a zombie creator's PID). An identity whose objects remain held past a 300-second bound stays billed, with an operator-visible warning, rather than releasing quota the kernel still accounts; the cost falls on the offending tenant's own budget. This design is evaluated in \S\ref{sec:gc-evaluation}.

\textbf{Dynamic policy management.}
Administrators modify quotas at runtime through the \texttt{tenant\_manager\_client} tool (\texttt{create}, \texttt{update}, \texttt{delete}, \texttt{query}); the daemon updates the Policy Segment in place, taking effect for new allocations without application restart. Existing allocations are not retroactively revoked, respecting the non-preemptable nature of RDMA resources.

\subsection{Security Analysis}
\label{sec:security-analysis}

The Interception Layer runs at the tenant's own privilege level, so a tenant can skip or overwrite checks that live inside its own address space (\S\ref{sec:intro}). Table~\ref{tab:coverage} summarizes which layer constrains each remaining path and what guarantee results; \S\ref{sec:eval-bypass} exercises each row.

\textbf{Creation paths.} In the normal case, an enrolled process calls a CP verb through \texttt{libibverbs}, and the interposer validates identity, checks quota in shared memory, and forwards approved calls. Deviating from that path, requests reach the RNIC through three kinds. All RDMA control commands (uverbs, \texttt{rdma\_cm}, umad) are ioctl or write operations on a device file descriptor, obtainable only through \texttt{open()}, whose entry the deployment guard controls, so unenrolled processes fail closed here. Calls issued through \texttt{libibverbs} inside an enrolled process pass the per-call wrappers and receive per-verb quotas (\S\ref{sec:interposition}). A tenant that legally opened the device but then issues commands outside the library---via \texttt{dlsym}, provider-specific direct verbs, or hand-packed uverbs commands---evades the wrappers; for such commands the guarantee degrades from per-type to aggregate: rdma-cgroup caps, which sit above the tenant's privilege, bound what they can commit, and periodic ledger--kernel reconciliation detects the resulting drift (\S\ref{sec:eval-bypass} measures all three outside-library forms). Per-type quotas therefore hold on the library path, and the aggregate cap covers the rest.

\textbf{Descriptor passing.} \texttt{SCM\_RIGHTS} escapes the guard as well: objects created on a received descriptor bill the \emph{receiver's} cap, so a tenant can drain a peer's quota by passing a descriptor and keeping the context alive---bounded by the victim's own cap, with connection-manager rate limiting as the standard mitigation.

\textbf{Ledger tampering.} The usage ledger stays writable at equal privilege, so we bound the consequences of forgery rather than the writes; the bound has three parts. In magnitude, rdma-cgroup caps sit above the tenant's privilege and bound what any forgery can commit. In time, reconciliation turns drift into a detected, self-healed divergence---about one second in our tests (\S\ref{sec:eval-bypass})---and since a determined tenant can rewrite the ledger after each heal, the bound holds per interval rather than permanently. In attribution, detection identifies divergence, not the writer: it shows \emph{that} the ledger disagrees with the kernel, not \emph{who} made it disagree. Two classes of state are excluded by construction: quotas and identity bindings live in daemon-owned, read-only segments (\S\ref{sec:accounting}), because a writable quota field would be a one-shot permanent denial that reconciliation cannot tell from an administrator's change.

\textbf{Ledger availability.} A corrupted ledger lock fails closed---admissions stall rather than exceed quotas---until a daemon restart resets the segment; the stall is loud and recoverable but not attributable, so an enrolled tenant can repeat it (hardening: owner-tracked locks with bounded spinning). Reconciliation then restores kernel truth (18.5~ms to the first admitted creation, about one second to full consistency in our tests).

\textbf{Fork whitening.} A forked child inherits the parent's device descriptor, so releasing accounting at the parent's death would under-report while kernel objects persist. The pending state of \S\ref{sec:gc-design} prevents this; without it, orphaned kernel QPs accumulated to 17 across ten fork generations in our test until the aggregate cap stopped creation---the cap bounding exactly the divergence that per-process accounting alone cannot prevent.

\begin{table}[t]
\centering
\caption{Attack-surface coverage: constraining layer, guarantee, and evaluation per path.}
\label{tab:coverage}
\footnotesize
\setlength{\tabcolsep}{2.5pt}
\renewcommand{\arraystretch}{0.9}
\begin{tabular}{@{}p{2.8cm}p{1.9cm}p{2.4cm}l@{}}
\toprule
\textbf{Attack path} & \textbf{Constraining layer} & \textbf{Guarantee} & \textbf{Eval.} \\
\midrule
Library path & interposition + quotas & per-type admission & \S\ref{sec:eval-exhaustion} \\
Unenrolled process & device guard & fail closed & \S\ref{sec:eval-bypass} \\
Enrolled, non-library calls & rdma-cgroup cap & aggregate bound & \S\ref{sec:eval-bypass} \\
Usage-counter forgery & reconciliation & detect + self-heal & \S\ref{sec:eval-bypass} \\
Descriptor passing & receiver quota & bounded by victim cap & \S\ref{sec:security-analysis} \\
Ledger-lock corruption & fail closed + restart & stall, recoverable & \S\ref{sec:security-analysis} \\
Fork whitening & pending state + cap & billed until release & \S\ref{sec:gc-evaluation} \\
Identity spoofing & kernel credential & billed to real tenant & \S\ref{sec:eval-bypass} \\
Unbound identity & registration & fail closed & \S\ref{sec:eval-bypass} \\
\bottomrule
\end{tabular}
\end{table}

\section{Implementation}
\label{Implementation}

Our implementation of the three-layer architecture comprises approximately 6,400 lines of C.
At load time, it opens \texttt{libibverbs.so} via \texttt{dlopen} and resolves real symbols via \texttt{dlsym}; deployment uses dynamic linker interposition~\cite{linkers}.

The library preserves exact IB Verbs API signatures, so approved calls reach the original provider functions and rejected calls return \texttt{EPERM} before any kernel invocation. The hardest parts are not the forwarding but the accounting around it: counters update with atomic increments under contention from mutually untrusting processes, quota reads must never observe a half-written policy entry (the sequence-number check in \S\ref{sec:accounting}), and identity must be established once at startup---the library registers over a Unix socket with kernel-provided credentials, after which \texttt{RDMA\_TENANT\_ID} is not an authorization input. All hot-path state lives in daemon-initialized POSIX shared memory (\texttt{shm\_open})~\cite{posix} organized into a header plus per-tenant and per-process accounting regions, synchronized by in-memory spinlocks rather than system calls. The daemon handles tenant creation, quota configuration, hot reconfiguration, and identity registration (\S\ref{Design}); it populates the client-mapped, read-only Binding Table at registration time.

\textbf{PID reuse detection.} Shared memory indexes resources by PID, so the layer reads each process's start time from \texttt{/proc/[pid]/stat} on every update and resets counters on mismatch (\S\ref{sec:gc-design}).

\textbf{Daemon crash resilience.} Shared memory persists independently of the daemon, so enforcement continues if the daemon crashes; upon restart the daemon re-attaches via \texttt{shm\_open} and resumes, with no state reconstruction.

\section{Evaluation}
\label{Evaluation}

We evaluate four questions: whether quotas prevent cache-exhaustion interference, whether an unmodified application benefits, what admission costs on the control plane and the data plane, and how accounting behaves under scale, crashes, and bypasses. All experiments use commercial RNICs and unmodified operating-system kernels; native RDMA provides the baseline where applicable.

\subsection{Experimental Setup}
We conduct tests on servers equipped with Mellanox ConnectX-6, connected via 100\,Gbps RDMA over Converged Ethernet version 2 (RoCEv2) links. The servers run standard Linux distributions without any kernel modules or firmware patches. We use standard RDMA performance testing tools from the perftest suite, including \texttt{ib\_write\_bw}, \texttt{ib\_read\_bw}, and \texttt{ib\_send\_bw}, all running without source code modification by loading the interception library through environment variables. \emph{Attack workloads} (QP-cache exhaustion and MR-cache flooding) are generated using the open-source Husky benchmark suite from Kong et al.~\cite{understandrdma}.

The topology consists of two servers directly connected via an RDMA network, each running tenant management daemons and the applications under test. Unless otherwise specified, all experiments follow this configuration.

\subsection{Resource Exhaustion Prevention}
\label{sec:eval-exhaustion}

This experiment validates \Name{}'s protection against resource exhaustion, configuring per-process QP quotas of 10, 50, 100, and 200 and MR count quotas of 50, 300, 600, and 1000; for each value the test program requests twice the quota, so any over-allocation would be exposed.

\begin{figure}[htbp]
    \centering
    \includegraphics[width=\linewidth]{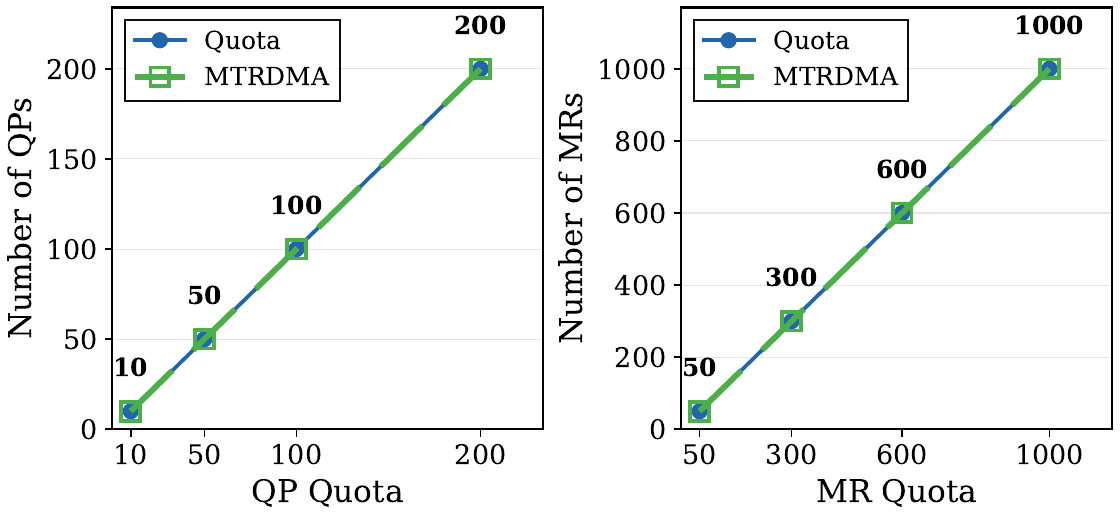}
    \caption{QP and MR quota enforcement accuracy: configured quota and created count overlap ($y=x$), zero deviation at all scales.}
    \label{fig:exp10_scalability}
\end{figure}

As shown in Figure~\ref{fig:exp10_scalability}, in all cases \Name{} creates exactly the number of QPs allowed by the quota and rejects all surplus requests---e.g., when the quota is 100, only 100 of the 200 requested QPs succeed and the remaining 100 are blocked.
Across all scales, \Name{}'s actual created MR count exactly matches the quota (Figure~\ref{fig:exp10_scalability}, right panel).

\textbf{Analysis.}
\Name{} enforces quotas with zero deviation for both QP and MR resources across a wide dynamic range: under sequential creation, actual usage matches the configured quota exactly; under concurrent creation, the per-process cache admits only in-flight creations within one TTL window, the observed peak never exceeded the quota in 40/40 runs at up to 128 racing threads, and a 3.2M-operation create/destroy churn across 32 threads closed with an exact zero ledger.

\subsection{QP Cache Attack Protection}
\label{sec:QPC-cache-attacks}

We replay the attack of \S\ref{sec:cache_attacks} under four schemes: the victim holds 91.5\,Gbps over 4 QPs while the unprotected noisy neighbor drives 100 QPs (5 instances of \texttt{ib\_send\_bw}, 20 QPs each) on the same physical RNIC port: (1)~native, (2)~\Name{}, (3)~PeRF~\cite{PeRF}, and (4)~Justitia~\cite{Justitia} (both deployed unmodified per their papers).

\begin{figure}[htbp]
\centering
\includegraphics[width=0.8\linewidth]{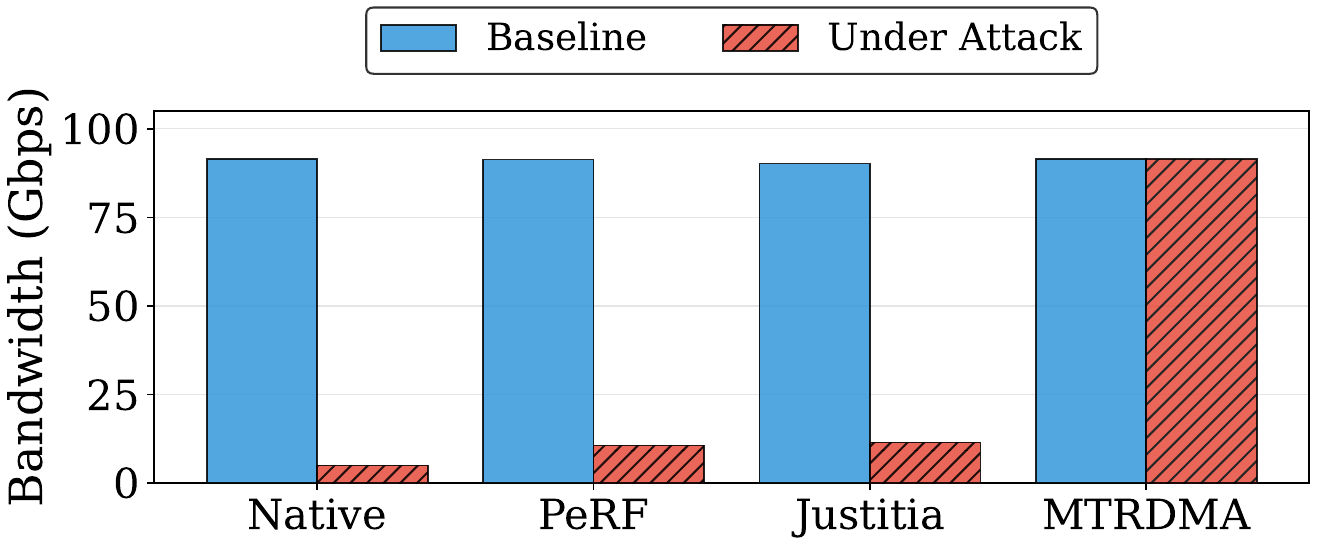}
\caption{QP Cache attack protection across four schemes (PeRF/Justitia: DP schedulers, no CP mediation)}
\label{fig:qpc_attack_comparison}
\end{figure}

Figure~\ref{fig:qpc_attack_comparison} compares victim bandwidth across the four schemes. Under the native configuration, the victim achieves a stable baseline of 91.52~Gbps; once the noisy neighbor activates its 100 QPs, bandwidth collapses to 5.00~Gbps (94.5\% degradation). With \Name{}, the victim retains 91.45~Gbps under attack (0.08\%); PeRF degrades 91.31$\rightarrow$10.63~Gbps (88.4\%) and Justitia 90.22$\rightarrow$11.45~Gbps (87.3\%): both schedule traffic from existing QPs but leave creation unmediated.

\textbf{Analysis.}
\Name{}'s pre-creation interception eliminates this degradation at the source; bandwidth-centric mechanisms remain ineffective because they do not control creation---the root cause of cache contention. DP scheduling and CP isolation are complementary.

\subsection{Generalization to MR Cache Exhaustion}
\label{sec:mr-attack}

\begin{figure}[htbp]
\centering
\begin{minipage}[t]{0.48\linewidth}
\centering
\includegraphics[width=\linewidth]{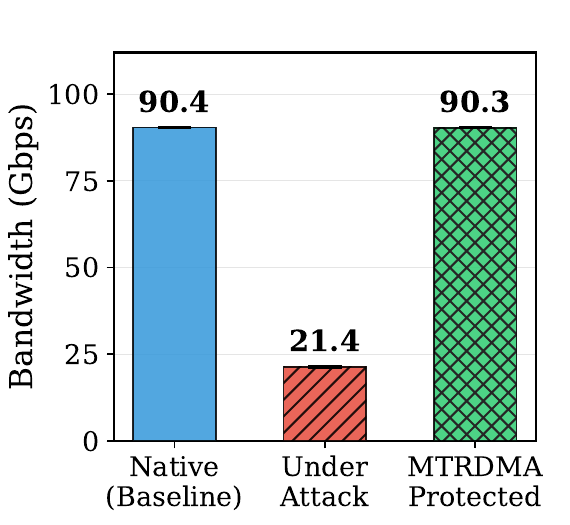}
\end{minipage}
\hfill
\begin{minipage}[t]{0.48\linewidth}
\centering
\includegraphics[width=\linewidth]{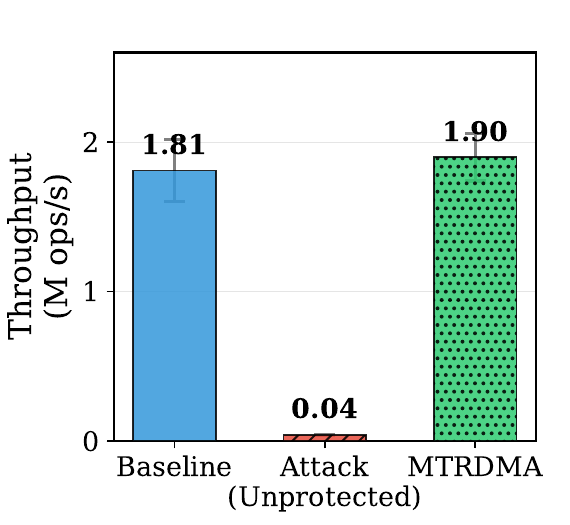}
\end{minipage}
\caption{Left: MR cache attack protection (victim bandwidth). Right: HERD throughput under baseline, unprotected attack, and \Name{} protection.}
\label{fig:app-protection}
\end{figure}

MR registration also consumes firmware-managed MPT/MTT entries~\cite{understandrdma}, and the same admission path covers this vector. With the victim running \texttt{ib\_write\_bw} (64~KB messages, one QP), a noisy neighbor registering 16{,}384 small MRs (512~bytes, four \texttt{RdmaEngine} instances) collapses the victim from 90.40 to 21.40~Gbps (76.3\% degradation); \Name{}'s per-tenant MR quota of 64 restores 90.29~Gbps (0.1\%, Figure~\ref{fig:app-protection}, left)---one admission mechanism covers both on-chip exhaustion vectors; byte-level MR accounting is an orthogonal extension~\cite{lwn2026rdmacg}.

\subsection{Real-Application Protection}
\label{sec:herd-protection}

We further evaluate \Name{} on HERD~\cite{kvrdma}, an RDMA key-value store whose mixed RC/UD QP usage exercises production-grade patterns.

\textbf{Experimental setup.} The victim runs unmodified HERD (single server, single client) with 1~RC~QP for requests and 1~UD~QP for datagrams (2~QPs total), plus 1,000~MRs. The noisy neighbor launches 16~\texttt{ib\_send\_bw} instances of 20~QPs each (320 total, 512-byte messages); under \Name{} it is capped at 20~QPs (Tenant~2) and HERD runs as Tenant~1 (QP quota~16).

Figure~\ref{fig:app-protection}(right) shows HERD throughput across three scenarios. At baseline, HERD achieves $1.81 \pm 0.21$~M ops/s (mean~$\pm$~std over 3 runs). Under attack without protection, throughput collapses to $0.04$~M ops/s---a \textbf{97.8\% degradation}. With \Name{} (noisy neighbor capped at 20~QPs), HERD recovers to $1.90 \pm 0.16$~M ops/s; the baseline and protected intervals ($\pm1\sigma$) overlap, so the 5\% difference is within measurement variance.

\textbf{Analysis.} The noisy neighbor's 320~QPs evict HERD's connection state from the on-chip QP Cache, forcing expensive cache misses on every operation; \Name{} caps its QP count at the quota via library-level \texttt{EPERM}, without modifying HERD. The attack thus reaches unmodified production applications, and one QP quota covers all QP types.

\textbf{Stock cgroup caps, measured.}
\S\ref{sec:existing_limits} sets the stock \texttt{rdma} cgroup cap up as the first of the two standard-path design choices; we now measure how far it actually gets us. Under the 100-QP flood from \S\ref{sec:QPC-cache-attacks} and an object cap of 20, the attacker sustains only eight QPs, and further clients are rejected. The victim's bandwidth falls by about 30\%, to 64.99~Gb/s, while the attack flows receive 25.35~Gb/s. Together they deliver 90.34~Gb/s of the 92.79~Gb/s baseline. The near-preserved aggregate indicates bandwidth contention, not cache thrashing: the victim's loss matches what the surviving attack flows consume.

The cap nevertheless provides only coarse containment, not a usable quota: it combines QPs, MRs, CQs, and PDs in one object budget. With the object cap set to 20 to contain the QP-cache attack, a tenant running HERD with two QPs can create only 13 4-KiB MRs---far short of the 1{,}000 MRs the application requires. Raising the cap to 1{,}000 or more to accommodate HERD would also give a tenant enough headroom to create 100 QPs and launch the same cache attack. We therefore use the stock cgroup cap as a coarse backstop for bounding out-of-model tenants (\S\ref{sec:security}), not as a substitute for per-resource quotas.

\subsection{Data-Plane Transparency}
\label{sec:eval-transparency}

We measure bandwidth across ten message sizes (512\,B--8\,MiB, four QPs per flow) and single-QP write latency on a 100\,Gb/s RoCEv2 link, three repetitions per scenario.

\begin{figure}[!htb]
    \centering
    \includegraphics[width=0.95\linewidth]{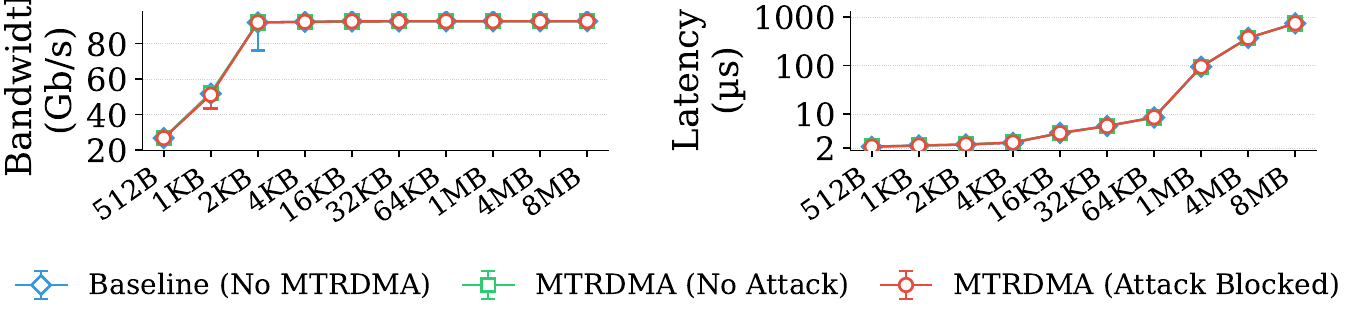}
    \caption{Data-plane transparency versus message size (100\,Gb/s RoCEv2). Left: bandwidth (four QPs per flow); right: single-QP write latency (log scale). Curves: native baseline, \Name{} without attack, and \Name{} with attack QP requests rejected by admission. Medians of 3 runs (10 for select baseline sizes); error bars span min--max.}
    \label{fig:exp6_bandwidth}
\end{figure}

\textbf{Results.}
Baseline medians (Figure~\ref{fig:exp6_bandwidth}, left) rise from 26.8~Gbps at 512\,B to 51.8~Gbps at 1\,KiB---a per-flow posting-rate ceiling of ${\sim}$6.2~Mmsg/s---and reach ${\sim}$92.6~Gbps from 2\,KiB upward, 1\,MiB included (no dip at four QPs). \Name{} without attack tracks the baseline within 1\% at every size, so interception alone leaves the data path untouched; under attack, with the attacker's QP requests refused by admission, the victim still matches no-attack within 2\% at every size. Median single-QP write latency (right) stays within ${\sim}$3\% of baseline in all three scenarios.

\textbf{CPU Utilization.}
We measure per-process CPU utilization during sustained \texttt{ib\_write\_bw} transfers (1~MB messages, 4~QPs, 30~seconds, 3 runs): the client process consumes $104.29 \pm 0.10$\% CPU natively vs.\ $104.43 \pm 0.00$\% with \Name{}---a negligible increase of 0.14\%.

\subsection{High-Density Tenant Scalability}
\label{sec:eval-scalability}

We configure a per-tenant QP quota of 10 and measure at 64, 100, 128, 192, and 256 concurrent tenant processes---whether the CP handles hundreds of tenants without degradation or unfairness---each continuously creating and destroying one QP for 10 seconds. Occupancy stays bounded by the quota (at most 2{,}560 QPs) while the cumulative load yields 253{,}141 creation attempts on one ConnectX-6 NIC; we measure success and failure counts, aggregate throughput with the Jain fairness index, and creation latency (P50, P99).

\begin{figure}[htbp]
\centering
\includegraphics[width=0.8\linewidth]{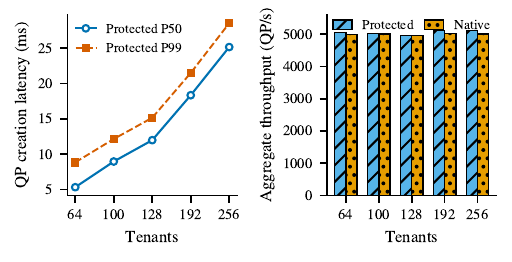}
\caption{High-density tenant scalability: (left) QP creation latency (P50, P99); (right) aggregate throughput and Jain fairness index, 64--256 tenants.}
\label{fig:high-density}
\end{figure}

Figure~\ref{fig:high-density}(left) shows P50 creation latency rising from 5.31~ms at 64 tenants to 25.16~ms at 256 (P99 from 8.83 to 28.53~ms), reflecting contention for the shared creation path as density quadruples. Over the same range, aggregate throughput stays flat at 4{,}970--5{,}120~QP/s ($\le$2\% variation), the Jain fairness index never drops below 0.999, and all 253{,}141 attempts succeed (Figure~\ref{fig:high-density}, right). Without \Name{}'s interception layer the identical workload shows the same growth, attributable to contention for the kernel's serialised QP-creation path rather than \Name{}'s accounting.

\subsection{CP Overhead}
\label{sec:eval-overhead}

We evaluate the CP overhead of each isolation scheme---QP creation and MR registration latency, which directly impact application startup---over 5 independent runs:
(i)~Baseline, standard RDMA with no isolation;
(ii)~\Name{}, with intercept library and resource quota checking;
(iii)~PeRF~\cite{PeRF}, with its modified \texttt{libmlx5}; and
(iv)~Justitia~\cite{Justitia}, with its rate-controlled environment.

\begin{figure}[htbp]
    \centering
    \includegraphics[width=0.82\linewidth]{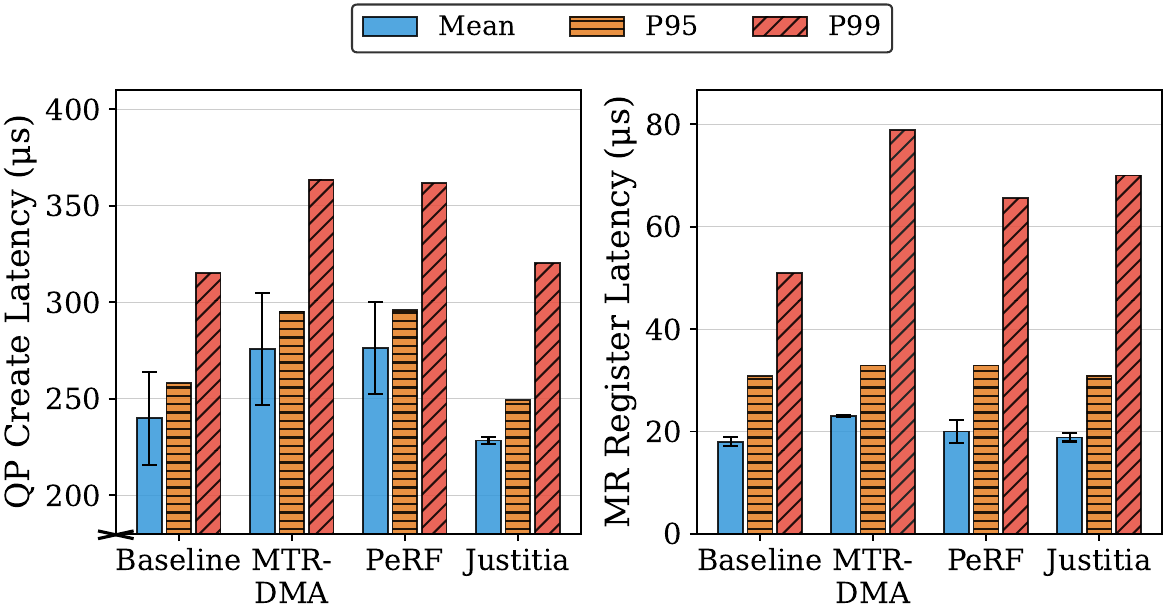}
    \caption{QP creation and MR registration latency across four isolation schemes (ConnectX-6). Mean, P95, and P99 over 5 runs of 1,000 iterations.}
    \label{fig:exp1_overhead}
\end{figure}

Figure~\ref{fig:exp1_overhead} reports QP creation and MR registration latency for the four schemes. QP creation averages $239.9 \pm 24.1~\mu\text{s}$ at baseline, with \Name{} ($275.8 \pm 29.2~\mu\text{s}$), PeRF ($276.5 \pm 23.8~\mu\text{s}$), and Justitia ($228.4 \pm 1.9~\mu\text{s}$) in the same range; MR registration means span $18.0$--$23.0~\mu\text{s}$ across all four. P99 tails also align ($315$--$364~\mu\text{s}$ for QP creation, $51$--$79~\mu\text{s}$ for MR registration).

\textbf{Analysis.}
PeRF's creation hook only records context and Justitia controls data-plane rates from a separate Pacer process, so neither adds meaningful CP cost; for \Name{}, the dominant cost ($>$95\%) remains the kernel ioctl round-trip for QP allocation and MR registration, which user-space interception cannot affect.

\subsection{GC Mechanism Evaluation}
\label{sec:gc-evaluation}

We evaluate reclamation after process termination: a test worker registers QP/MR resources and is terminated via various signals---graceful \texttt{exit()}, \texttt{SIGKILL}, \texttt{SIGTERM}, and five workers receiving \texttt{SIGKILL} simultaneously.

\begin{figure}[t]
    \centering
    \includegraphics[width=0.8\linewidth]{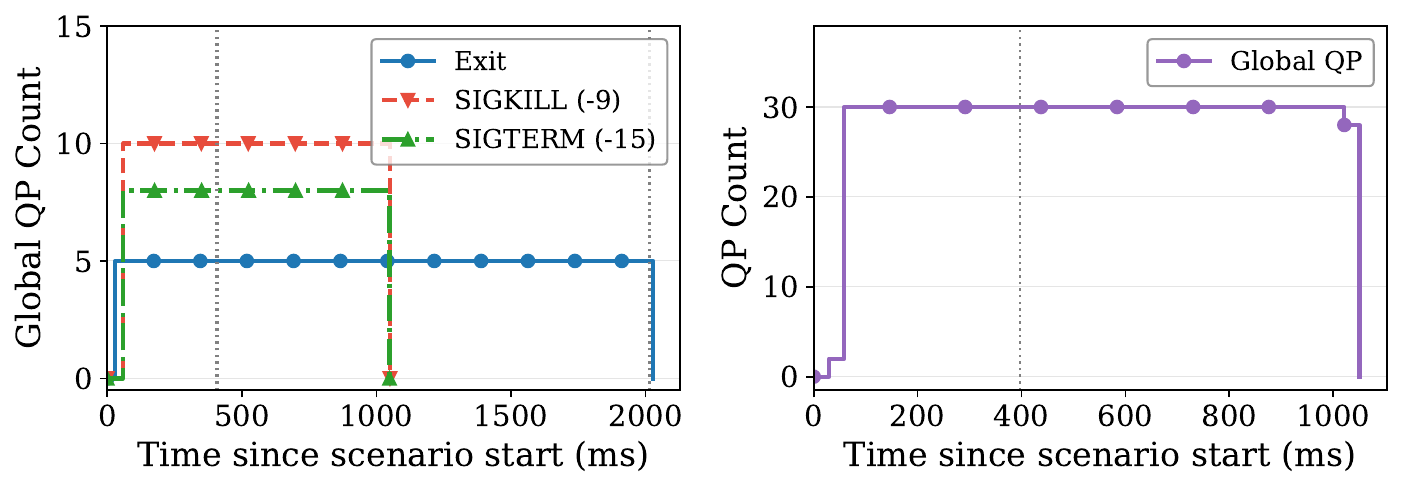}
    \caption{GC reclamation timelines, rebased to scenario start (dotted line: signal delivery). Left: \texttt{exit()} vs.\ signals. Right: concurrent five-worker \texttt{SIGKILL}; global QP count.}
    \label{fig:gc-reclamation}
\end{figure}

Figure~\ref{fig:gc-reclamation} shows the reclamation timelines, rebased to each scenario's start; in all scenarios the ledger and the global count return to zero with every allocated QP accounted for---complete reclamation without leakage. Graceful \texttt{exit()} reclaims within 14~ms through the library's own destruction path. \texttt{SIGKILL}, \texttt{SIGTERM}, and the concurrent five-worker storm are reclaimed by the daemon's event-driven \texttt{pidfd} collector within 640--653~ms of termination, the concurrent case confirming that the periodic fallback catches events lost under high-concurrency exits.

\textbf{Fork retention.}
When a parent process forks a child and then exits, the child inherits the parent's ledger entry, and quota is reclaimed only after the child exits as well. In our experiment, a parent creates 10 QPs, forks a child, and exits; the ledger entry transfers to the child, and once the child exits the quota is reclaimed within 1.05~s. Processes of the same tenant share one ledger entry: with a quota of 12 and 10 consumed, a sibling process attempting to create 3 QPs succeeds in creating only 2. The complementary failure mode---release at the parent's death while kernel objects persist---is bounded by the aggregate cap (\S\ref{sec:security-analysis}).

\subsection{Closing the Direct-Device Bypass}
\label{sec:eval-bypass}

\begin{figure}[htbp]
\centering
\includegraphics[width=0.85\linewidth]{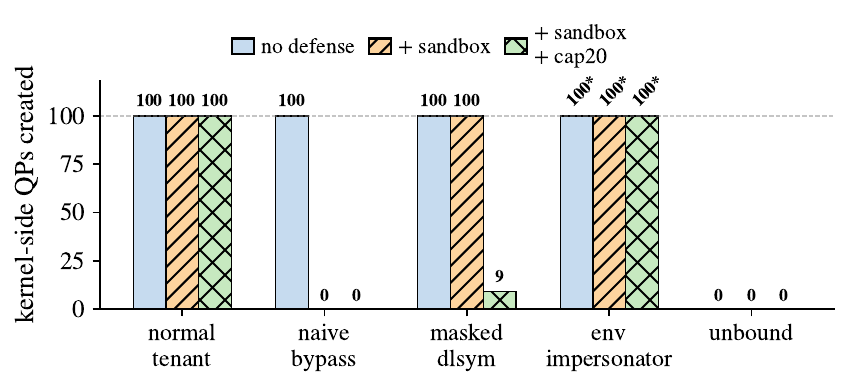}
\caption{Layered defense and identity binding, measured as kernel-side QPs created. Roles request 100~QPs under escalating defense: the sandbox stops the naive bypass; the cap bounds the masked tenant; the env impersonator is billed to its own tenant ($^{*}$: victim ledger $+0$); the unbound role fails at registration.}
\label{fig:defense-matrix}
\end{figure}

\Name{}'s quotas are enforced at the interception layer, so a tenant that bypassed the library would evade them; \S\ref{sec:security-analysis} states each path's bound, and this section measures it. A \emph{naive} attacker never loads the interception library (static linking and direct syscalls reduce to this form) and targets the sandbox's guarantee that a process outside \Name{}'s runtime obtains no RDMA; a \emph{masked} tenant keeps the library mapped but resolves the real symbols via \texttt{dlsym}. An enrolled tenant can also bypass from within, via provider direct verbs or hand-issued raw \texttt{RDMA\_VERBS\_IOCTL} commands. We evaluate these paths as five roles (Figure~\ref{fig:defense-matrix}), each requesting 100~QPs.

\textit{Layer 1 (sandbox).} The naive attacker succeeds without defense (100/100, ledger 0) but is stopped by the sandbox alone: 0/100, \texttt{EPERM} at device open.
\textit{Layer 2 (object cap).} The masked tenant passes the sandbox by design---the library is mapped---and creates 100/100~QPs without defense; the rdma-cgroup object cap (20 objects) bounds it to 9, every further attempt failing \texttt{EAGAIN}. The cap cannot express per-type quotas: $\sim$11 per-context driver objects leave room for only 9~QPs (\S\ref{sec:security}).
\textit{Layer 3 (reconciliation).} In both successful bypasses the ledger records zero while the kernel confirms the QPs; the daemon's periodic ledger--kernel reconciliation catches this divergence, self-healing every drift within 1.06~s across 45 tamper trials.
\textit{Direct-verbs and raw forms.} An enrolled tenant calling \texttt{mlx5dv\_create\_qp} created 19 of 20 QPs (the twentieth hit the aggregate object cap, \texttt{EAGAIN}) with the ledger still at zero; reconciliation self-healed it to 19 in 1.07~s and to 0 in 0.97~s after cleanup. Raw hand-issued ioctls created 20/20 QPs; in a cap-boundary run 25 of 30 succeeded before the cap stopped the rest, reconciliation converged in 1.17~s, and the ledger zeroed in 0.95~s after exit. One raw-created QP also completed an end-to-end SEND (27-byte payload), so outside-library commands reach beyond object creation (a functional check, not a benchmark).

In the identity cases, an impersonator claiming another tenant's ID via the environment variable is billed to its own tenant (the claimed tenant's ledger does not move); a process in the victim's cgroup setting \texttt{RDMA\_TENANT\_ID} is billed to the victim; and a fresh cgroup without a binding-table entry is denied 20/20 at registration, before reaching the kernel.

\textbf{Binding-table enforcement.}
The daemon-owned Binding Table is client-mapped read-only (\S\ref{sec:accounting}): escalating the mapping to writable (\texttt{mprotect}) or re-opening the object read-write fails with \texttt{EACCES} for the tenant but succeeds for a same-UID control, confirming root ownership as the protection source.

\textbf{Griefing the writable ledger.}
Three attacks target the tenant-writable usage ledger directly (\S\ref{sec:threatmodel}). \textit{Usage-counter forgery}: overwriting a victim's \texttt{qp\_count} with its quota denied the next creation in 4 of 5 trials; healing took at most 1.04~s (\S\ref{sec:security-analysis}). \textit{Quota forgery}: overwriting a victim's quota with zero denied its creations indefinitely (reconciliation cannot distinguish an administrator's change from a forgery, \S\ref{sec:accounting}); with quotas in the daemon-owned, read-only Policy Segment, re-opening it read-write is refused with \texttt{EACCES}. \textit{Global-counter forgery}: overwriting the global QP count denied every subsequent creation (5/5) and persisted until manual reset; the daemon now rebuilds the global snapshot from verified per-tenant usage each reconciliation, so the same attack self-heals within 1.05~s.

\section{Discussion}
\label{Discussion}

\textbf{Operational hardening and deployment.}
\label{sec:security}
\Name{} operates under the threat model of \S\ref{sec:threatmodel}; deployment reduces to three steps. \emph{Enrollment:} run the Resource Management Daemon under an operator-controlled host identity, install the operator-owned cgroup-to-tenant rules, and enroll each workload with the interception library (preload). \emph{Device guard:} attach the access guard as an OCI seccomp-notify policy so unenrolled processes obtain no RDMA device (enrollment is self-enforcing). \emph{Caps:} place each workload in its SLA-derived rdma-cgroup (self-migration fails with \texttt{EACCES}), bounding even library-bypassing tenants.

\textbf{Known limitation and hardening path.} The sandbox's device-open decision admits a known TOCTOU race~\cite{seccomp_unotify}, closed structurally by the kernel-documented \texttt{SECCOMP\_IOCTL\_NOTIF\_ADDFD} mitigation (supervisor-side open plus fd injection), which our guard supports. Two smaller surfaces remain: registration-socket flooding (standard rate limiting applies) and inbound connections that burn a victim's own quota. \Name{} cannot \emph{preempt} allocated resources (hardware non-preemptability); operators terminate the offender; the hybrid GC reclaims the orphans within about one second of kernel release (\S\ref{sec:gc-evaluation}). A forked child can hostage its parent's quota past a 300-second bound (\S\ref{sec:gc-design}); the cost falls on the offending tenant's own budget.

\textbf{Hardware scope and future work.}
\label{sec:hardware-qos}
Validation uses ConnectX-6 (100~Gbps RoCEv2) with supplementary App.~F BlueField-3 host-mode results; interception targets the vendor-agnostic \texttt{libibverbs} ABI. Hardware QoS~\cite{TIMELY,HPCC,Swift_CC} bounds DP rates and complements \Name{}'s CP quotas. Cross-vendor validation (Intel E810~\cite{iRDMA}, Broadcom), DPU-offload mode, and iWARP (same ABI, unvalidated) remain future work. QP limits can be derived empirically (HERD 2--4 QPs; MPI one per peer); automatic tuning of quotas remains future work as well.

\section{Related Work}
\label{RelatedWork}

\textbf{Security-oriented RDMA mechanisms.} sRDMA~\cite{sRDMA} adds NIC-based authentication and encryption, and ReDMArk~\cite{ReDMArk} characterizes RDMA security and availability attacks; neither enforces per-tenant control-plane quotas on commodity RNICs. \textbf{RDMA virtualization and connectivity.} SR-IOV~\cite{SR-IOV} exposes per-tenant virtual functions without performance isolation~\cite{understandrdma}. Software alternatives~\cite{FreeFlow,MasQ,LITE,RCN} mediate QP creation for connectivity and address virtualization; quota accounting, identity binding, and crash-safe reclamation are outside their design, and their daemons fail the admission-latency bound (\S\ref{sec:existing_limits}).

\textbf{Microarchitecture contention characterization.} Husky~\cite{understandrdma} and Collie~\cite{Collie} characterized cache-exhaustion attacks without deployable isolation; concurrent work mitigates them on BlueField-3 containers via telemetry-driven throttling~\cite{kim2025noisy}. \Name{} targets the same threat class with creation-time admission.

\section{Conclusion}
\label{Conclusion}
\Name{} closes the control-plane resource-isolation gap in multi-tenant RDMA through lightweight user-space interception. Hierarchical quotas on QP and MR creation---enforced without touching the data path or modifying kernels, drivers, firmware, or applications---cut victim degradation under cache-exhaustion attacks from 94.5\% to 0.08\%, and the system scales to 256 tenants. Within the managed-runtime threat model of \S\ref{sec:threatmodel}, \Name{} bounds what a misbehaving tenant can commit, detects what it can distort, and reclaims what it leaves behind.

\appendix

\appendix
\section{Complementary Role with DP Schedulers}
\label{sec:complex-attacks}

We evaluate \Name{}+PeRF against simultaneous DP bandwidth contention and CP QPCache exhaustion. The victim (4 QPs, 1 MB) achieves $\sim$91.5 Gbps baseline. The DP noisy neighbor uses 16 QPs (1 MB) to saturate the link, while the CP noisy neighbor spawns 100 QPs (512 B) to overflow the QPCache. All three roles use distinct \Name{} tenant IDs with independent QP quotas (victim: 16, DP: 64, CP: 16), and PeRF governs all tenants.

\begin{figure}[t]
\centering
\includegraphics[width=0.65\linewidth]{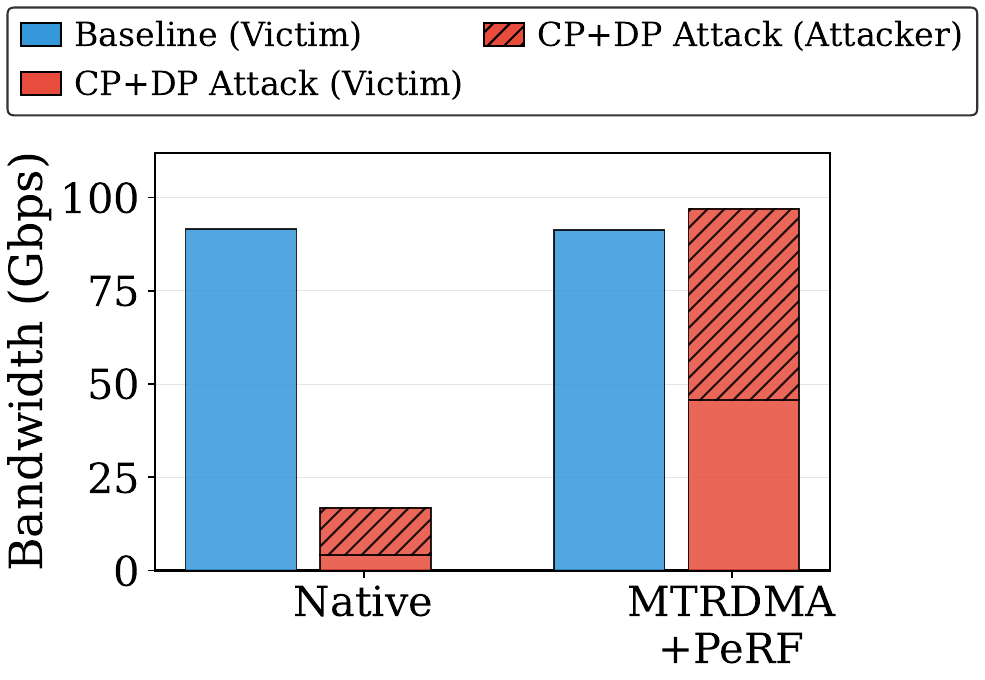}
\caption{\Name{}+PeRF under simultaneous CP+DP attacks: CP isolation restores the victim's fair share of DP bandwidth. (left) Without protection, the victim bandwidth falls to \textbf{4.2~Gbps}; the noisy neighbor captures the remainder, yielding only 16.8~Gbps total. (right) Under protection, the victim maintains \textbf{45.7~Gbps}---its fair share of the bandwidth under three-way DP contention---with near-line-rate total throughput (97.0~Gbps). The stacked bars show victim (red, bottom) and noisy neighbor (red, top) bandwidth shares.}
\label{fig:combined}
\end{figure}

Figure~\ref{fig:combined} presents the results. Under native configuration, the combined CP+DP attack reduces the victim to 4.2 Gbps with only 16.8 Gbps total throughput---both CP and DP are compromised. With \Name{}+PeRF, the victim sustains 45.7 Gbps under the combined attack with 97.0 Gbps total throughput---an 11.0$\times$ improvement over native.

We note that the victim's 45.7 Gbps is approximately half of its 91.5 Gbps baseline. This is expected: \Name{} fully neutralizes the CP attack by capping the noisy neighbor's QP creation at 16 (preventing QPCache pollution), but PeRF (a DP bandwidth scheduler) fairly divides the 100 Gbps link among three tenants contending for bandwidth. The victim receives approximately its fair share (100~Gbps $\div$ 3~tenants $\approx$ 33~Gbps, with the remaining variance due to message size differences and PeRF's weighted scheduling). Thus, the 45.7~Gbps does not indicate residual CP interference---it reflects the DP bandwidth competition that \Name{} is not designed to address. The near-line-rate total throughput (97.0~Gbps) confirms the QPCache is fully protected; no bandwidth is wasted to cache misses.

These results validate that \Name{}'s CP isolation is orthogonal to DP scheduling. When co-deployed with existing DP schedulers, the combined stack sustains near-line-rate aggregate throughput under simultaneous attacks, confirming CP--DP separation as a viable architecture. The fair-sharing limitation is inherent to any CP-only isolation mechanism and motivates the need for complementary DP schedulers.

\section{WQE Cache Attack Defense}
\label{sec:wqe_attack}

Beyond QP and MR exhaustion (\S5.3 and \S5.4), RNIC microarchitectural caches also include WQE (Work Queue Entry) state. This experiment validates \Name{}'s protection against WQE cache attacks that exhaust the on-chip Recv WQE cache. The victim runs \texttt{ib\_write\_bw} with 64 KB messages while the noisy neighbor runs four \texttt{RdmaEngine}\cite{understandrdma} instances, each creating 8 QPs with 1,024 outstanding receive Work Requests (WRs) per QP, totaling 32 QPs and 32,768 Recv WRs; under \Name{} the global QP count is capped at 24.

\textbf{Implementation Insight:} We initially attempted to limit the attack by capping \texttt{max\_recv\_wr} (the QP's Recv queue depth) in addition to QP count. However, our validation revealed that the kernel driver internally rounds up user-requested WR depths via \texttt{roundup\_pow\_of\_two()} in \texttt{set\_rq\_size()}, and on our test platform the firmware enforces a minimum effective depth of 4,096, effectively ignoring user-space limits below this threshold. Consequently, WR-depth-only limiting proved ineffective on this hardware, and the defense relies entirely on QP count enforcement.

\begin{figure}[ht]
    \centering
    \includegraphics[width=0.65\linewidth]{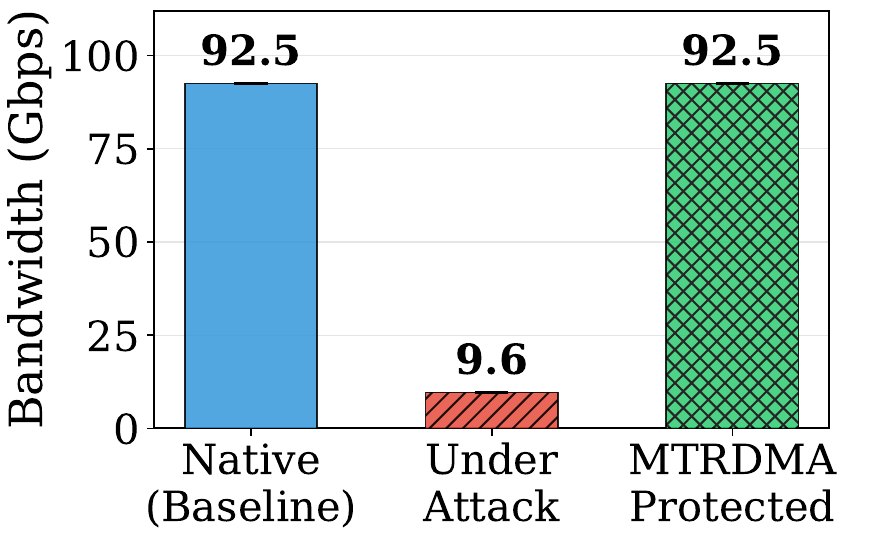}
    \caption{WQE cache attack protection via QP quota enforcement.}
    \label{fig:app_wqe_attack}
\end{figure}

Figure~\ref{fig:app_wqe_attack} shows the victim achieves 92.55~Gbps at baseline, collapses to 9.61~Gbps (89.6\% degradation) when unprotected, and recovers to 92.55~Gbps (0\% degradation) with \Name{}. These results confirm that WQE cache attacks are a critical multi-tenant threat. However, the protection is achieved indirectly: the kernel driver rounds up \texttt{max\_recv\_wr} via \texttt{roundup\_pow\_of\_two()}, and on our test platform the firmware enforces a minimum effective depth of 4,096, so capping QP count also bounds the total WQE footprint. This does not constitute independent WQE-level isolation; a noisy neighbor who stays within the QP quota could still consume disproportionate WQE cache capacity by configuring deep WR queues.

\textbf{WQE footprint bound under QP quotas.} Although the defense is indirect, QP quotas impose a hard upper bound on a tenant's aggregate receive-WQE footprint. A tenant holding at most $Q$ QPs with per-queue depths $d_i$ occupies $\sum_{i \leq Q} d_i$ receive-WQE entries. On our platform each receive queue occupies at least 4,096 entries after driver rounding, so the evaluated workloads exhibit a footprint of $Q \times 4{,}096$ entries; in general the footprint is bounded by $Q \times d_{\max}$, where $d_{\max}$ is the maximum per-queue depth a tenant can request. Tightening $Q$ therefore linearly shrinks a tenant's worst-case WQE-cache pressure without affecting already-established connections. Fully independent WQE-level enforcement---bounding not only creation-time depth but also runtime queue occupancy---would still require driver support (§6.3); the depth-level admission check below closes the creation-time half of this gap inside the interception layer.

\begin{figure}[ht]
    \centering
    \includegraphics[width=0.9\linewidth]{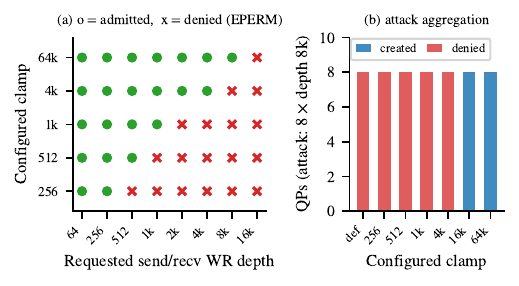}
    \caption{Depth-level admission enforcement: (a) requested WR depth vs.\ configured clamp (green = admitted, red = denied with \texttt{EPERM}); (b) aggregate deep-queue attack (8 QPs $\times$ depth 8{,}192) under each clamp.}
    \label{fig:e2-wqe-clamp}
\end{figure}

\textbf{Depth-level admission enforcement.}
\Name{} additionally implements the depth half of this bound directly at the interception layer: a configurable clamp (\texttt{MAX\_SEND\_WR\_LIMIT}/\texttt{MAX\_RECV\_WR\_LIMIT}) rejects any \texttt{ibv\_create\_qp()} whose requested send or receive depth exceeds it, failing with \texttt{EPERM} before the driver commits hardware state. Figure~\ref{fig:e2-wqe-clamp}(a) sweeps the requested depth against five clamp configurations: the admission boundary tracks the clamp exactly (default 1{,}024), and our NIC further caps the grantable depth at 8{,}192. Because the firmware rounds receive queues up to the 4{,}096-entry floor described above, clamps below 4{,}096 do not shrink the per-queue footprint further; the clamp instead bounds depth \emph{above} the floor, capping a tenant's worst-case footprint at $Q \times \max(4{,}096, C)$ entries for clamp $C$. Figure~\ref{fig:e2-wqe-clamp}(b) demonstrates the bound under an aggregate deep-queue attack---eight QPs each requesting a depth of 8{,}192: an unclamped tenant pins 8~MiB of send/receive WQE buffers, whereas every clamp $\le$4{,}096 denies all eight requests, reducing the attack to the $Q$-bounded case already covered by the QP quota.

\textbf{Why fast admission and reclamation matter despite long QP lifetimes.} QPs are long-lived in steady state, so creation rate is not a steady-state concern; however, creation arrives in bursts during job startup, scale-out, and failure recovery. For example, an $N$-rank all-to-all MPI job creates $N(N{-}1)$ QPs in one startup burst, and a replaced rank must re-establish its connections before the job can proceed. A synchronous, kernel-mediated admission path would add per-operation latency precisely during these bursts and prolong recovery. \Name{}'s fast path adds only $\sim$36\,$\mu$s over the $\sim$240\,$\mu$s kernel path and sustains ${\sim}5{,}000$ QP/s under high tenant density (§5.7, §5.8), keeping admission off the burst-time critical path. Conversely, long lifetimes make \emph{reclamation} the binding requirement: after a crash, stale quota can block a replacement workload from creating new QPs and increase recovery latency, which is exactly the leak the hybrid GC reclaims within 0.7\,s of termination (§5.9).

\textbf{Quota boundary semantics.} Quota enforcement is intentionally abrupt rather than gradual: within quota, established connections run at full line rate; at the boundary, new creations fail explicitly with \texttt{EPERM}---an error applications can detect, log, and retry---rather than the silent cache-eviction degradation the mechanism prevents. Under-provisioning therefore surfaces as admission failure instead of performance decay. For elastic or connection-heavy workloads, however, under-provisioning can block scale-out or rank replacement, so operators should provision quotas with demand-based headroom (§6.3); e.g., HERD requires only 2 QPs and sees no rejections at per-tenant quotas of 16--20 (§5.5).

\section{Dynamic Policy Hot Update}
\label{sec:eval-hotupdate}

We evaluate online quota adjustment without application restart. A tenant with an initial QP quota of 6 attempts to create QPs at a fixed interval; the administrator first lowers the quota to 5 \emph{in place}---the 6 existing QPs are retained, respecting non-preemptability---and then raises it to 12. Updates use an in-place \texttt{UPDATE\_QUOTA} operation that preserves the tenant and its allocations, and moves the reconciliation baseline (\S5.10) so that a legitimate policy change is not misread as ledger drift.

\begin{figure}[ht]
    \centering
    \includegraphics[width=\linewidth]{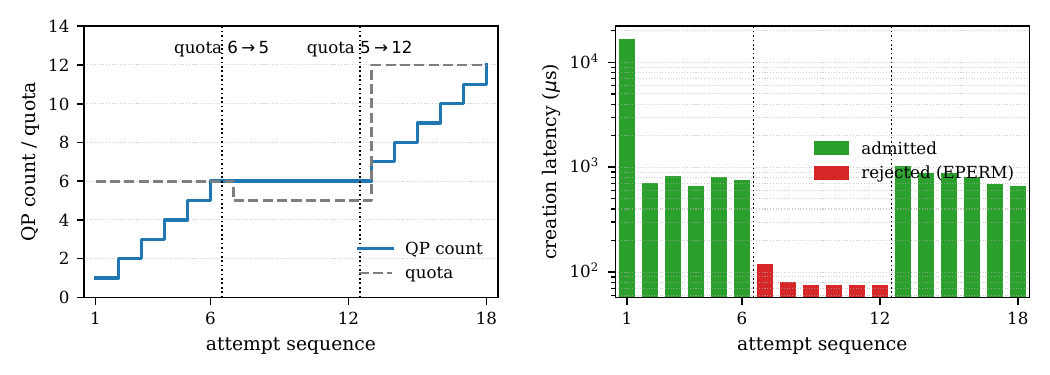}
    \caption{Dynamic policy hot-update with in-place quota changes ($6 \rightarrow 5 \rightarrow 12$; dotted lines mark the updates). Left: QP count (solid) versus quota (dashed). Right: per-attempt creation latency (green = admitted, red = rejected).}
    \label{fig:exp5}
\end{figure}

Figure~\ref{fig:exp5} shows the 18 recorded attempts: 6/6 succeed at quota 6; after the downward update the existing QPs remain but new creations fail (0/6, \texttt{EPERM}); after the upward update creations succeed again (6/6). The daemon applies each update in ${\sim}$2.5\,ms, and the first attempt after the update confirmation (${\sim}$150\,$\mu$s later) already reflects the new quota---no application restart, and no effect on established QPs.

\section{Cache Effectiveness Evaluation}
\label{sec:eval-cache}

We compare quota-check latency between direct shared-memory access and the per-process local cache (256-bucket hash table, 100~ms TTL). Under all tested workloads, cache hit rates reach 100\%.

\begin{figure}[ht]
    \centering
    \includegraphics[width=\linewidth]{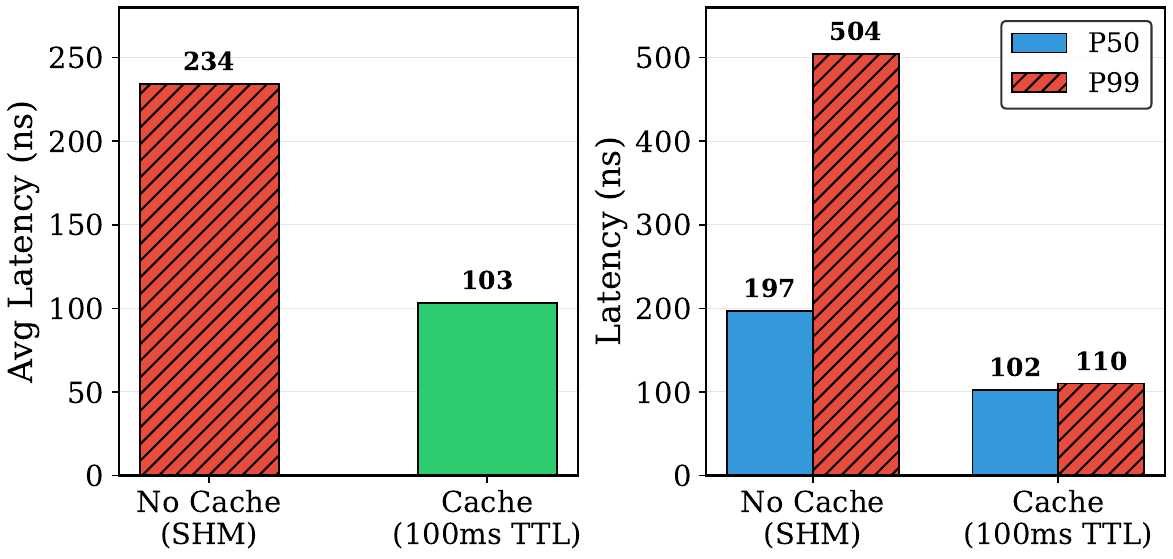}
    \caption{Cache effectiveness: latency comparison between shared memory and local cache.}
    \label{fig:exp4_cache}
\end{figure}

Without caching, average latency is 234~ns (P99: 504~ns). With caching, average latency drops to 103~ns (P99: 110~ns), achieving a 2.27$\times$ speedup and 78\% P99 reduction. Figure~\ref{fig:exp4_cache} illustrates this improvement. The adaptive TTL strategy shows negligible hit-rate difference ($<$0.03\%) versus fixed TTL under stable workloads.

\textbf{Daemon-arbiter baseline.}
To quantify the speed leg of the Challenge~1 trilemma (\S1), we benchmark the three candidate quota-check paths, 300k operations each: the per-process optimistic cache averages ${\sim}$0.13\,$\mu$s, a direct shared-memory read ${\sim}$0.23\,$\mu$s, and a daemon arbitration round trip over a Unix socket 20.0\,$\mu$s (20--35\,$\mu$s across runs)---over two orders of magnitude slower than the local cache, and serialized through a single process whose unavailability would block every tenant's admission. This baseline also confirms that the $\sim$5{,}000 QP/s creation throughput in \S5.7 reflects the kernel's ioctl path, not the admission layer.

One might instead close the check-then-act window by pre-deducting quota with an atomic fetch-and-add before delegating to the real verb, refunding on failure. We evaluated this design and retained check-then-act with the optimistic cache: pre-deduction adds a refund atomic on the failure path; the transient per-tenant overshoot it would eliminate is already bounded, as the authoritative ledger plus the global hard cap prevent over-allocation (\S3.2); and as a resource-containment mechanism (\S3.1) \Name{} tolerates the millisecond-bounded staleness. Moreover, the measured gap between the cache path (${\sim}$0.13\,$\mu$s) and a direct shared-memory check (${\sim}$0.23\,$\mu$s) is four orders of magnitude below the kernel's $\sim$240\,$\mu$s creation path (\S5.7), so pre-deduction buys no measurable admission latency.

\section{Hierarchical Quota Enforcement}
\label{sec:hierarchical-quota}

We validate \Name{}'s three-level quota enforcement: per-process, per-tenant, and global. A single process attempts to create 5--20 QPs against process-level caps of 1, 3, 5, 8, and 16. Tenant-level quotas of 4, 6, and 10 are enforced via the Tenant Manager Daemon. Global quotas are tested by having two processes share a node-wide budget.

\begin{figure}[ht]
    \centering
    \includegraphics[width=0.7\linewidth]{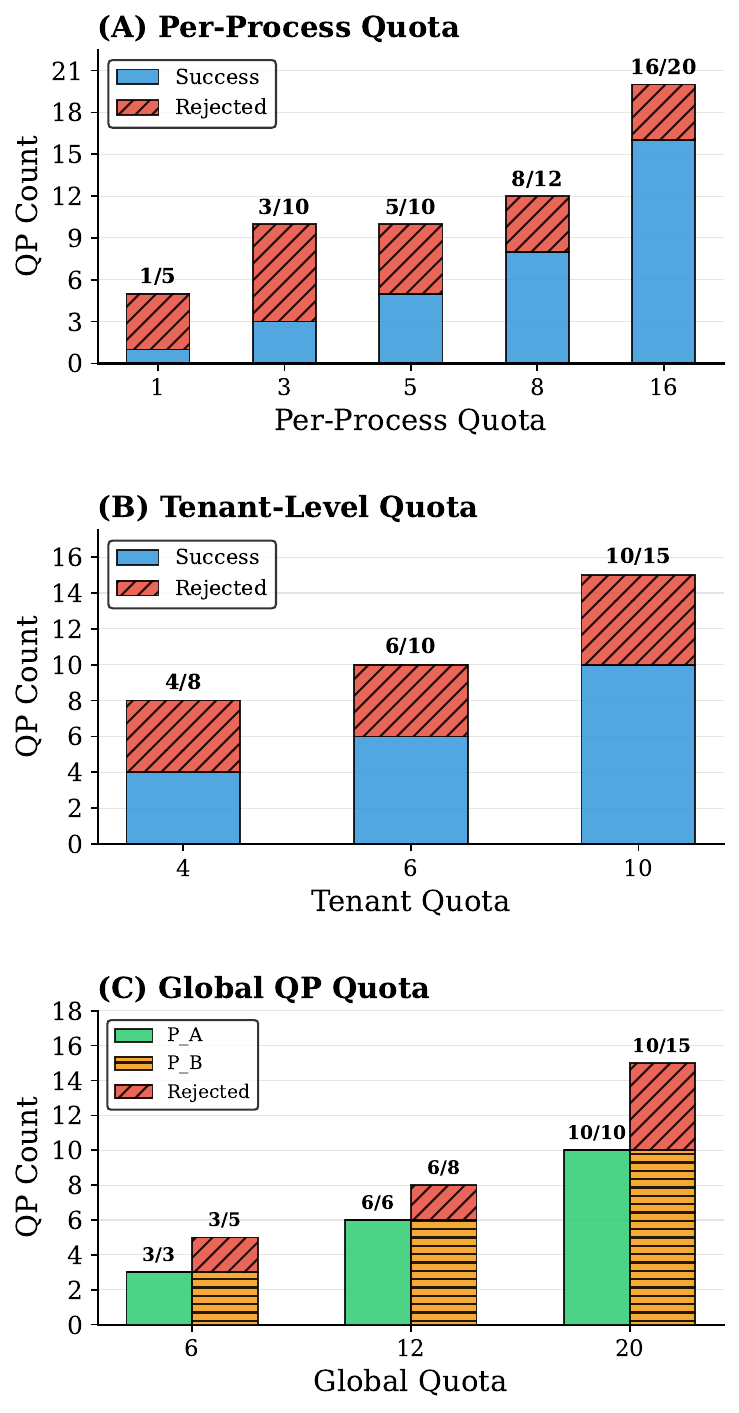}
    \caption{Hierarchical quota enforcement. (A)~Per-process quota. (B)~Tenant-level quota. (C)~Global QP quota shared across processes.}
    \label{fig:hierarchical-quota}
\end{figure}

Figure~\ref{fig:hierarchical-quota} shows that all layers independently and precisely cap resource creation. Per-process success counts exactly match the configured caps (1/5, 3/10, 5/10, 8/12, 16/20). Tenant-level quotas enforce hard ceilings (4/8, 6/10, 10/15). Global quotas are correctly shared: when the limit is 6, two processes each obtain 3 QPs; when the limit is 12, the first acquires 6 and the second is capped at the remaining 6. When multiple layers are active simultaneously, the effective limit is the minimum of the active quotas, with \texttt{EPERM} returned immediately upon exhaustion of any layer.

\section{Host-Mode Validation on BlueField-3}
\label{sec:bf3}

We validate \Name{} on NVIDIA BlueField-3 in \textbf{host-mode (NIC) operation}, where RDMA verbs execute on the host CPU via the standard mlx5/\texttt{libibverbs} path---architecturally identical to ConnectX-6/7. The testbed pairs the BlueField-3 (mlx5\_0, host-side \texttt{libibverbs}) with a ConnectX-6 DX peer over a 50~Gbps rate-limited link. The victim runs \texttt{ib\_write\_bw} (1~MB, 4~QPs); the noisy neighbor launches 5 \texttt{ib\_send\_bw} instances (20~QPs each). Table~\ref{tab:bf3} summarizes the protection gradient across four configurations.

\begin{table}[h]
\centering
\small
\setlength{\tabcolsep}{3pt}
\caption{BlueField-3 validation: victim bandwidth under increasing noisy neighbor QP quotas. The 50~Gbps rate-limited link achieves 43--46~Gbps baseline, confirming near-line-rate conditions.}
\begin{tabular}{@{}lccc@{}}
\toprule
\textbf{Quota} & \textbf{Baseline(Gbps)} & \textbf{Under Attack(Gbps)} & \textbf{Degradation} \\
\midrule
N/A & 45.73 & 1.71 & 96.3\% \\
100 & 45.77 & 3.02 & 93.4\% \\
8 & 43.40 & 37.39 & 13.8\% \\
4 & 45.74 & 45.76 & 0\% \\
\bottomrule
\end{tabular}
\label{tab:bf3}
\end{table}

The gradient demonstrates precise dose-response: as the noisy neighbor's QP quota tightens from 100 to 4, victim degradation drops from 93.4\% to 0\%. At quota = 4, all QP creation attempts beyond the limit are rejected with \texttt{EPERM}, leaving the QPCache intact and the victim unaffected. This confirms that \Name{}'s library-level interception is compatible across two generations of Mellanox/NVIDIA RNICs (ConnectX-6 and BlueField-3 host-mode). \textbf{Data Processing Unit (DPU)-offload mode, where verbs execute on the embedded Advanced RISC Machine (ARM) cores through a different software path and ABI, is not covered by the current host-mode design} and remains future work alongside cross-vendor validation.

\clearpage 

\bibliographystyle{ACM-Reference-Format}
\bibliography{reference}

\end{document}